\documentclass[11pt,a4paper]{article}
\pdfoutput=1
\usepackage[pdftex]{graphics}
\usepackage{jheppub}
\usepackage{amsmath,amssymb,amsfonts}
\usepackage{enumitem}
\usepackage{multirow}
\usepackage{array,booktabs}
\usepackage{slashed}

\newcommand{\be}{\begin{eqnarray}}
\newcommand{\ee}{\end{eqnarray}}
\newcommand{\eqa}{\begin{eqnarray}}
\newcommand{\eqae}{\end{eqnarray}}

\newcommand{\bn}{\begin{enumerate}}
\newcommand{\en}{\end{enumerate}}
\newcommand{\bl}{\begin{align}}
\newcommand{\el}{\end{align}}

\def\e{\epsilon}
\def\ve{\varepsilon}

\def\k{\kappa}

\def\m{\mu}
\def\n{\nu}

\def\r{\rho}

\def\s{\sigma}

\def\w{\omega}
\def\tr{{\rm tr}}

\def\jmath{{j}}
\def\bl#1\el{\begin{align} #1 \end{align}}
\def\bg#1\eg{\begin{gather} #1 \end{gather}}

\def\bld#1\eld{\begin{aligned} #1 \end{aligned}}
\def\bgd#1\egd{\begin{gathered} #1 \end{gathered}}

\newcommand{\eqc}[1]{(\ref{#1})}

\title{Complete Hamiltonian for spinning binary systems at first post-Minkowskian order}

\author[1]{Ming-Zhi Chung} 
\author[1,2]{Yu-tin Huang}
\author[3]{Jung-Wook Kim}
\author[3,4,5]{Sangmin Lee} 

\affiliation[1]{Department of Physics and Astronomy, National Taiwan University, Taipei 10617, Taiwan}
\affiliation[2]{Physics Division, National Center for Theoretical Sciences, National Tsing-Hua University, No.101, Section 2, Kuang-Fu Road, Hsinchu, Taiwan}
\affiliation[3]{Department of Physics and Astronomy, Seoul National University, Seoul 08826, Korea}
\affiliation[4]{Center for Theoretical Physics, Seoul National University, Seoul 08826, Korea}
\affiliation[5]{College of Liberal Studies, Seoul National University, Seoul 08826, Korea}

\emailAdd{dchung0741@gmail.com}
\emailAdd{yutinyt@gmail.com}
\emailAdd{jwkonline@snu.ac.kr}
\emailAdd{sangmin@snu.ac.kr}

\abstract{
Building upon recent progress in applying on-shell amplitude techniques to classical observables in general relativity, we propose a closed-form formula for the conservative Hamiltonian of a spinning binary system at the 1st post-Minkowskian (1PM) order.
It is applicable for general compact spinning bodies with arbitrary spin multipole moments. The formula is linear in gravitational constant by definition, but exact to all orders in momentum and spin expansions. At each spin order, our formula implies that the spin-dependence and momentum dependence factorize almost completely. 
We expand our formula in momentum and compare the terms with 1PM parts of the post-Newtonian computations in the literature. Up to canonical transformations, our results agree perfectly with all previous ones. 
We also compare our formula for black hole to that derived from a spinning test-body near a Kerr black hole via the effective one-body mapping, and find perfect agreement. 
}

\begin{document}
\begin{flushright}
\vspace{10pt} \hfill{NCTS-TH/2004} \vspace{20mm}
\end{flushright}

\maketitle

\newpage
\section{Introduction}
The on-shell approach, characterized as exploiting kinematic constraints combined with unitarity and symmetry principles to bootstrap physical observables, has often uncovered unexpected structures hidden in conventional formalism. Indeed this has been one of the highlights in the study of scattering amplitudes for the past decade. On the other hand, as it was demonstrated long ago, (electromagnetic and) gravitational two-body potentials can be extracted from singular limits of scattering amplitudes~\cite{Donoghue:1996mt, BjerrumBohr:2002kt, Khriplovich:2002bt,Holstein:2008sw,Holstein:2008sx,Holstein:2008sy}, naturally one would expect that the hidden structures found in scattering amplitudes will leave its fingerprint~\cite{Neill:2013wsa, Bjerrum-Bohr:2013bxa}. Indeed structures such as the double copy~\cite{Kawai:1985xq} and color-kinematic duality~\cite{Bern:2008qj}, has already found its foothold in simplifying the computation of classical potentials~\cite{Bjerrum-Bohr:2014zsa, Bjerrum-Bohr:2016hpa, Shen:2018ebu, Bern:2019nnu}. See \cite{Bern:2019crd} for the most up to date results and review.

Concurrently, massive spinor-helicity formalism introduced by one of the authors~\cite{Arkani-Hamed:2017jhn}, has exposed the hidden simplicity of the multipole moments of spinning black holes, where the infinite number of Wilson coefficients (set to unity for Kerr black holes) can be captured by a single on-shell three-point matrix element~\cite{Chung:2018kqs,Chung:2019duq,Aoude:2020onz}.
This simplicity has led to a streamlined computation of spin effects in the scattering angle~\cite{Guevara:2018wpp}, linear and angular impulse~\cite{Guevara:2019fsj} of rotating black holes, as well as new insights into the origin of shift relations between rotating and Schwarzschild black hole solutions~\cite{Arkani-Hamed:2019ymq}.

For systems with a well defined separation of scales (the size of the object, the orbital radius and the wavelength of radiation), the dynamics of the compact objects coupled to gravity can be approximated by a worldline action~\cite{Goldberger:2004jt}. For compact spinning objects, the worldline theory attains extra spin fields~\cite{Porto:2005ac}, which introduce an infinite number of multipole moments each with its own Wilson coefficient~\cite{Porto:2008jj, Levi:2015msa}. In~\cite{Chung:2019duq}, on-shell three-point amplitude was derived for such worldline actions with general Wilson coefficients. This provides the residue of the single graviton exchange between two spinning bodies in the limit which the transfer momentum $q^\mu$ is small, $q^2\rightarrow 0$, which yields the long range dynamics at leading order in $G$.  Importantly, special care was required to account for the extra spin effects 
due to the fact that the little group space of distinct particles are related by Lorentz boosts. This Thomas precession factor, termed  “Hilbert space matching” in~\cite{Chung:2019duq}, was computed at the leading Post Newtonian (PN) order, leading to the spin-dependent part of the conservative Hamiltonian to all order in spins, but leading order in PN for each spin degree. The validity of the result was confirmed by matching to the all order in spin result for rotating black holes when the Wilson coefficients are set to unity~\cite{Vines:2017hyw}. 

In this paper, we find the exact form of the Thomas precession factor, given as:
\begin{align}
U^{(a)} = \exp\left[ - i \left(\frac{m_b}{r_a E}\right) \ve(q,u_a,u_b,a_a) \right] \,,
\quad 
r_a \equiv 1 + \frac{E_a }{m_a}  \,, 
\quad
E = E_a + E_b \,.
\end{align}
where $a,b$ label the two bodies, $u_{a,b}$ their proper velocities and $a_{a,b}$ their spin vectors normalized by their respective masses. Equipped with this precession factor, the general form of the two body potential at the 1st post-Minkowskian order (1PM) can be expressed as:
\begin{align}
V_{\rm 1PM}^{\rm (general)} = - \frac{4\pi G m_a^2 m_b^2}{E_a E_b} \int \frac{d^3 \vec{q}}{(2\pi)^3} e^{i\vec{q}\cdot \vec{r}}
\left[\frac{1}{2} \sum_{s=\pm 1} e^{2s\theta} W_a(s\tau_a) W_b(s\tau_b)\right] U^{(a)} U^{(b)} \,.
\label{V-1PM-master-preview}
\end{align}
where $W_{a,b}$ are the generating functions for the Wilson coefficients for the compact spinning bodies, 
and $\tau_{a,b}$ are some Lorentz scalars proportional to the spin length vectors $a_{a,b}$. 
The precise definitions will be given in section~\ref{sec:amp}.
When all the Wilson coefficients are set to 1, the general potential specializes to the black hole potential as:
\begin{align}
V_{\rm 1PM}^{\rm (BH)} = - \frac{G m_a^2m_b^2}{2E_a E_b} 
\sum_{s=\pm 1} 
e^{2s\theta}
\left| \vec{r} + s\frac{E (\vec{p}\times \vec{a}_0)}{m_a m_b \sinh\theta} - \frac{\vec{p}\times \vec{a}_a}{m_a r_a} - \frac{\vec{p}\times \vec{a}_b}{m_b r_b} 
\right|^{-1} \,. 
\label{V-BH-final-preview}
\end{align}
where $\cosh\theta = u_a \cdot u_b$, $\vec{a}_0=\vec{a}_a + \vec{a}_b$, and $\vec{p}$ is the center of mass momentum. 

By expanding the potential \eqref{V-1PM-master-preview} in $|\vec{p}|^2$, including the precession factors $U^{(a)} U^{(b)}$, we obtain the Hamiltonian to arbitrary orders in PN expansion. As a consistency check, for results that are already in the literature, we have verified that ours match nontrivially through appropriate canonical transformations, which we summarize in Table~\ref{PN-check-table}. 
We leave the comparison at the NLO cubic-in-spin terms (marked by a star in Table~\ref{PN-check-table}) for a future work, as the result in \cite{Levi:2019kgk} at its current form requires more than a canonical transformation to be compared with ours. 

\begin{table}[ht]
\begin{center}
\begin{tabular}{l|c|c|c|c}
 & LO & NLO & NNLO & N$^3$LO
\\
\hline
$S^1$ & \cite{Levi:2015msa}  & \cite{Levi:2015msa} & \cite{Levi:2015uxa}&  ?
\\
\hline
$S_a S_b$ &   \cite{Levi:2015msa} & \cite{Levi:2014sba} & \cite{Levi:2014sba} & - 
\\
\hline
$S_a^2$ &   \cite{Levi:2015msa} &  \cite{Levi:2015msa} &   \cite{Levi:2016ofk} & - 
\\
\hline
$S^3$ & \cite{Levi:2014gsa} & \cite{Levi:2019kgk}$^{\star}$ & - & - 
\\
\hline
$S^4$ & \cite{Levi:2014gsa} & - & - & - 
\end{tabular}
\caption{References we used to compare our results to 1PM parts of PN computations.}
\end{center}
\label{PN-check-table}
\end{table}

While equivalent up to canonical transformations, our results demonstrate further hidden simplicity. First, the on-shell approach naturally lands us on the so-called ``isotropic gauge",  where all  $(\hat{n}\cdot \vec{p})$ terms are missing ($\hat{n}$ being the unit vector in the radial direction of the two objects), echoing that in~\cite{Bern:2019crd}. 
Furthermore, we find that the spin dependent term factorizes almost completely from the $|\vec{p}|^2$ dependence. In other words, a term in the potential that is of degree $m ,n$ in spin-vectors $S_a$, $S_a$, respectively, can be schematically written as (for example with $m+n$ even)
\eqa
V_{S_a^m S_b^n}= \left(\frac{G m_a m_b }{r^{m+n+1}} \right) \left[ F_{(m,n)}(\vec{a}_a,\vec{a}_b,\hat{n}) \right] X_{(m,n)}(\vec{p}^2) \,, 
\eqae
where the function $F_{(m,n)}$ is independent of $|\vec{p}|^2$. Said in another way, in our representation $S\cdot p$ terms are absent. Indeed such terms are ubiquitous in the results that have been listed in the literature so far, yet we explicitly show that through a canonical transformation the former are in complete agreement with our result.  

In addition to reproducing known PN results (and producing new ones), which is perturbative 
both in spin and momentum, we perform a check valid to all orders in spin. In \cite{Vines:2017hyw}, 
Vines used the effective one-body (EOB) mapping \cite{Buonanno:1998gg,Damour:2016gwp,Damour:2017zjx}
to read off the 1PM potential 
between two Kerr black holes from the exact Kerr solution. Using well known EOB dictionary at the 1PM level, 
we will show that the potential of Vines is equivalent to ours shown in \eqref{V-BH-final-preview}.

This paper is organized as follows. 
In section~\ref{sec:amp}, we review the tree-level amplitude for general compact spinning bodies 
and how to dress it with the Thomas precession factor to derive the complete 1PM potential 
exact in spin and momentum. 
In section~\ref{sec:spin-order}, we expand the exact formula in spin, up to quartic order in spin 
while keeping exact dependence on momentum.
In section~\ref{sec:checks}, to compare with PN computations in the literature, we further expand the results of section~\ref{sec:spin-order} in momentum. 
We identify the explicit form of canonical transformations to match our results to the ones in the literature. The references we used for comparison are summarized in table~\eqref{PN-check-table}.
In section~\ref{sec:EOB}, specializing to Kerr black holes, we perform a check that is valid to all order in spin. We compare the exact potential \eqref{V-BH-final-preview} for black holes 
to the one obtained by Vines \cite{Vines:2017hyw} through an EOB mapping and find perfect agreement.
We conclude with some discussions in section~\ref{sec:conclusion}.

\section{Complete 1PM potential from amplitude} \label{sec:amp}

\begin{figure}
\begin{center}
\includegraphics[scale=0.5]{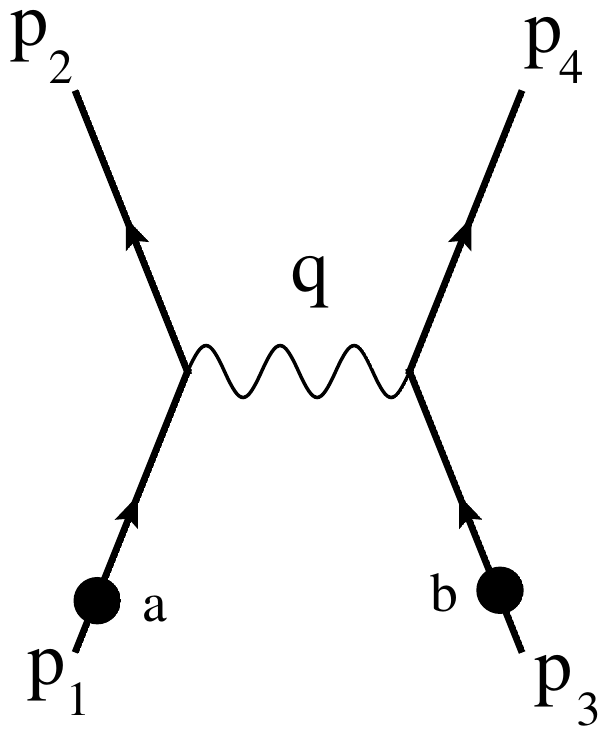}
\caption{The graviton exchange diagram between source $a$ and $b$ that yields the leading $1/q^2$ singularity, which is responsible for the classical potential.}
\label{GraviExchange}
\end{center}
\end{figure}

The 1PM classical potential can be extracted from the singular limit of a single graviton exchange between two compact spinning objects, i.e. the $2\rightarrow 2$ elastic scattering amplitude shown in Fig.~\ref{GraviExchange}. The kinematic set up in the center of mass frame is given by
\eqa\label{Kin1}
p_1 = ( E_a, \vec{p} {+} \vec{q} / 2 ), \;\;
p_3 = ( E_b, {-} \vec{p}{-} \vec{q} / 2 ), \;\;  
p_2 = ( E_a, \vec{p} {-} \vec{q} / 2 ) , \;\;
p_4 = ( E_b, {-} \vec{p} {+} \vec{q} / 2 ) \,,
\eqae
where the exchanged momentum $q^\mu=(p_{1}-p_2)^\mu=(0,\vec{q})$ is space-like. For the classical limit, we expand in small $|q|$, and since in Lorentzian signature this translate to the zero momentum limit, we will analytically continuing to complex (or split signature) momenta. In this case we can have $|q|\rightarrow0$ correspond to null momenta, $q^2=0$. The advantage of such analytic continuation is that with $q^2=0$, the amplitude factorizes into the product of two three-point amplitudes. This approach was introduced by Guevara~\cite{Guevara:2017csg} and named the \emph{holomorphic classical limit} (HCL). The leading order potential is extracted as 
\eqa
V(p,q)\sim\left.\frac{M_4(s,q^2)}{4 E_aE_b}\right|_{q^2\rightarrow0}\,.
\eqae 
We write $\sim$ here because for spinning objects, the particles are irreps in distinct little group space which are related via Lorentz boosts. These boosts will introduce additional spin-dependent factors which needs to be accounted for. The associated matching procedure was introduced in~\cite{Chung:2019duq}, termed \textit{Hilbert space matching}, and computed to leading PN order for each spin operator of given degree. This factor is the well known Thomas precession, for which we derive its exact form in this section, thus deriving the complete 1PM potential.

\subsection{The kinematics}
We begin with the kinematic setup for the single graviton exchange in the HCL limit. In terms of four-vectors, we have
\begin{align}
p_1 = p_a + q/2 \,, \quad p_2 = p_a - q/2 \,, \quad 
p_3 = p_b -q/2 \,, \quad p_4 = p_b + q/2 \,.
\label{kin-scattering}
\end{align} 
which in the COM frame reduces to eq.(\ref{Kin1}) . Asymptotically, the spinning particles are free, and characterized by 
their momenta and (Pauli-Lubanski) spin vectors. Rescaling them by the masses 
give the proper velocity and ``spin-length" vectors:
\begin{align}
u_\mu = \frac{1}{m} p_\mu \,, 
\quad 
a_\mu = \frac{1}{m} s_\mu \,.
\end{align}
We denote the Lorentz invariant amplitude by $\mathcal{M}$ and 
the non-relativistic one by $M$. The two are related by 
\begin{align}
M = \frac{1}{4E_a E_b} \mathcal{M} \,.
\end{align}
We adopt (with slight modification) the kinematic variables of by Bern et al.~\cite{Bern:2019nnu}:
\begin{align}\label{eq:Def}
\begin{split}
&m = m_a + m_b \,, \quad
\nu = m_a m_b /m^2 \,,  
\\
E_{a,b} = \sqrt{\vec{p}^2 + m_{a,b}^2} \,, \quad 
&E = E_a + E_b \,, \quad 
\xi = E_a E_b / E^2 \,, \quad 
\gamma = E/m \,,
\\
&\sigma = \frac{p_a \cdot p_b}{m_a m_b} = u_a \cdot u_b \equiv \cosh\theta \,.
\end{split}
\end{align}
The 1st and the 3rd lines are Lorentz invariant, whereas the 2nd line is specific 
to the COM frame. In the non-relativistic (NR) limit $\sigma\rightarrow 1$ and $\theta\rightarrow 0$. The analytically continued HCL kinematics is characterized in a Lorentz invariant way as 
\begin{align}
q^2 = p_a \cdot q = p_b \cdot q = 0\,. 
\end{align}
This implies, for example, 
\begin{align}
\pm i \ve_{\m\n\r\s} p_a^\m p_b^\n q^\r a^\s = m_a m_b (\sinh\theta) (q \cdot a) \,.
\label{pole-strange}
\end{align}
which can be derived by squaring both sides and identifying the determinant of the Gram matrix for the LHS. The sign ambiguity in the above can  also be seen from the definition of $\theta$ in eq.(\ref{eq:Def}), where it is invariant under $\theta\leftrightarrow -\theta$. As we will see later on, our potential will be an even function of $\theta$, and thus the ambiguity is irrelevant.\footnote{The difference for the two choices will be purely imaginary, and is relevant when considering electromagnetic interactions associated with dyons~\cite{Caron-Huot:2018ape} and gravitational dynamics in Taub-NUT space-time~\cite{Huang:2019cja}.} 
\subsection{1PM amplitude} 

The tree-level graviton exchange between two massive scalars is given as 
\begin{align}
\mathcal{M} &= - (16\pi G) \frac{m_a^2 m_b^2}{q^2} (2\sigma^2 - 1) 
= - (16\pi G) \frac{m_a^2 m_b^2}{q^2} \cosh(2\theta) \,.
\end{align}
The spinning analogue was computed by on-shell methods for Kerr black holes in \cite{Guevara:2018wpp,Chung:2018kqs,Guevara:2019fsj} and 
then generalized to general compact spinning bodies in \cite{Chung:2019duq}.
In our conventions, the result of \cite{Chung:2019duq} can be written as 
\begin{align}
\mathcal{M}_{\rm bare} =  - (16\pi G) \left(\frac{m_a^2 m_b^2}{q^2}  \right)
\left[ 
\frac{1}{2} \sum_{s=\pm 1} 
e^{2s\theta} W_a(s\tau_a) W_b(s\tau_b) \right] \,.
\label{spin-amp-bare}
\end{align}
We call it the bare amplitude to emphasize that it is missing a factor to be discussed later in this section. 
The (quantum) amplitude is a matrix element between asymptotic states, each of which is labelled by momentum and spin. Quantum spins take discrete values, but we are only interested 
in the classical limit where spin becomes effectively continuous. To take the classical limit including spin, one has to strip off systematically the polarization tensors of asymptotic states, 
as explained in \cite{Guevara:2018wpp,Chung:2018kqs,Guevara:2019fsj,Maybee:2019jus}. 
We do not intend to repeat the discussion of the classical limit here, except to note that 
the spin-dependent factors missing from \eqref{spin-amp-bare} are rooted in the polarization tensors. 

We should also stress that \eqref{spin-amp-bare} is only the leading piece of the amplitude in the $q^2\rightarrow 0$ limit, i.e. the ``leading singularity" of the exchange diagram, which will be sufficient to determine the 1PM potential. 

The variables $\tau_{a,b}$ in \eqref{spin-amp-bare} are defined by 
\begin{align}
\tau_{a,b} = i \frac{\ve(q,u_a,u_b,a_{a,b})}{\sinh\theta} \,,
\quad 
\ve(a,b,c,d) = \ve_{\m\n\r\s} a^\m b^\n c^\r d^\s \,,
\label{def:tau}
\end{align}
where the spin-length vectors $a_{a,b}^\m \equiv S_{a,b}^\m/m_{a,b}$ 
are regarded as classical variables, even though they originate from quantum operators 
during the computation of the amplitude. With eq.(\ref{def:tau}), we see that eq.(\ref{spin-amp-bare}) is an even function of $\theta$ as advertised.

Despite its appearance, $\tau_{a,b}$ is not singular in the NR limit $\theta \rightarrow 0$, since the invariant ``area" spanned by the two velocity vectors, $u_a$ and $u_b$, is precisely $\sinh \theta$. In other words, the anti-symmetric tensor,  
\begin{align}
\w_{\m\n} \equiv \frac{1}{\sinh\theta} \ve_{\m\n\r\s} u_a^\r u_b^\s \,,
\end{align}
reflects only the orientation of the 2-plane spanned by the two velocities. 

The functions $W_{a,b}$ encode the gravitational couplings. 
As discussed in~\cite{Chung:2019duq}, the infinite spin-dependent worldline operators that linearly couples to the Riemann tensor can be incorporated into a three-point amplitude of two spin-$s$ particles and one massless graviton, with the understanding that $s$ is to be taken to infinity, i.e. the classical spin limit. The three-point amplitude takes the form, 
\eqa\label{3ptAmp}
\mathcal{M}_{{\rm 3pt},s}^{2 \eta} =  \frac{\k m x^{2\eta} }{2} \ve_2^{\ast} \left[ \sum_{n=0}^{2s} \frac{C_{n}}{n!} \left( - \eta \frac{q \cdot S}{m} \right)^n \right] \ve_{1}  \,,
\label{eq:1bd3ptAmp}
\eqae
where $\ve$ represents the polarization tensors for the spin-$s$ particles, with integer spin, and $\eta = +1$ for positive helicity graviton and $\eta = -1$ for negative helicity graviton. The $x$ factor is a scalar function that carries the helicity weight of the massless graviton. Its explicit form will not be important here and can be found in~\cite{Arkani-Hamed:2017jhn}.

The general compact spinning body is characterized \cite{Levi:2015msa} by the Wilson coefficients   $C_{2n} = C_{{\rm ES}^{2n}}$ $(n\ge 1)$ and 
$C_{2n+1} = C_{{\rm BS}^{2n+1}}$ $(n\ge 1)$. It is convenient to include $C_0 = 1 = C_1$ and define the generating function: 
\begin{align}\label{GenDef1}
W(\tau) = \sum_{n=0}^\infty \frac{C_n}{n!} \tau^n \,.
\end{align}
For a Kerr black hole, $C_n = 1$ for all $n$ such that $W(\tau) = e^\tau$. 

It is sometimes useful to separate the even and odd parts of the generating functions, 
$W_{\pm} = \frac{1}{2}\left[W(\tau) \pm W(-\tau)\right]$,
so that we can write
\begin{align}
\begin{split}
\mathcal{M}_{\rm bare} &= {-} (16\pi G) \left(\frac{m_a^2 m_b^2}{q^2}  \right)\frac{1}{2} \sum_{s=\pm 1} 
e^{2s\theta} W_a(s\tau_a) W_b(s\tau_b) 
\\
&= {-}16\pi G\frac{m_a^2 m_b^2}{q^2}\left[ \cosh(2\theta) (W_{a+}W_{b+} {+} W_{a-}W_{b-}) {+} \sinh(2\theta) (W_{a+}W_{b-} {+} W_{a-}W_{b+})  \right]\,.
\end{split}
\label{spin-odd-even-x}
\end{align}
As noticed by \cite{Guevara:2018wpp,Chung:2018kqs,Guevara:2019fsj}, when both spinning bodies are Kerr black holes, the amplitude 
takes a particularly simple form: 
\begin{align}
\mathcal{M}_{\rm bare}^{\rm (BH)} =  - (16\pi G) \frac{m_a^2 m_b^2}{q^2}  
\cosh\left( 2\theta  + i \frac{\ve(q,u_a,u_b,a_0)}{\sinh\theta} \right) \,, 
\label{spin-amp-BH}
\end{align}
where $a_0^\m = a_a^\m + a_b^\m$ is the total spin-length vector. To extract the classical potential, the above result needs to be dressed by additional factors coming from definition of polarization tensors. Such factors have been referred to as \emph{Hilbert space matching} in~\cite{Chung:2019duq}.

\subsection{Thomas-Wigner rotation} \label{sec:rotation}
The amplitude by definition is a matrix element between distinct (little group) Hilbert spaces, one for each asymptotic state. The momentum for each asymptotic state serves as the reference to which the little group is defined. Since two asymptotic momenta for the same particle can be related via Lorentz boosts, the amplitude contains non-trivial rotation factors simply from the action of mapping between the distinct Hilbert space. To see this effect, let's setup a canonical little group frame for our two body problem. We begin with the reference momenta $p_0$ identified as the center of mass momenta, i.e. 
\bl
p_{0,a/b} &= \frac{m_{a/b}}{\sqrt{(p_1 {+} p_3)^2}} (p_1 {+} p_3) \,,
\label{eq:RefMomDef}
\el
where $p_{0,a/b}$ are appropriately normalized for particle $a,b$ respectively. This allows us to set up reference polarization vectors, which form the basis of the little group space. Let's focus on particle $a$, $b$ follows accordingly. Since $p_0$ is at rest, the polarization vector takes the form
\eqa
\e^{\mu}_i (p_0)=\delta^{\mu}_i\,.
\eqae
Now the polarization vector for generic momentum $p$ can be obtained by applying the boost that transforms $p_0$ to $p$, i.e. $G(p;p_0)^{\mu}\,_{\nu}$, and
\eqa
p = G(p;p_0) p_0 \,, \quad \e^\m(p) = G(p;p_0) \e (p_0) \,. \label{eq:BoostDef}
\eqae
Using this, we can relate the polarization vectors between in- and out-momenta, 
\eqa\label{Relate1}
\e (p_{\rm out}) &= G(p_{\rm out};p_0) G(p_{\rm in};p_0)^{-1} \e (p_{\rm in})= G(p_{\rm out};p_0) G(p_{0};p_{\rm in}) \e (p_{\rm in})\,.
\eqae
As a consequence even the simple contraction of two polarization vectors $\epsilon^*_{out}\cdot \epsilon_{\rm in}$, contains non-trivial spin factors:
\eqa
\e^{\ast \m} (p_{\rm out})\e_{\m} (p_{\rm in})&=&  \e^{\ast \m} (p_{\rm in}) \left[ G(p_{\rm out};p_0) G(p_{0};p_{\rm in})\e (p_{\rm in})\right]_\m\,,
\eqae
where the operator sandwiched between the two polarization vectors now acts on the basis vectors of the same little group space, i.e. $\e (p_{\rm in})$. Now the operator contains both boosts and rotations, and we are only interested in the latter part. Since the rotation leaves the momenta $p_{\rm in}$ unchanged, it can be identified with 
\eqa
G(p_{\rm in};p_{\rm out})G(p_{\rm out};p_0) G(p_{0};p_{\rm in})\,.
\eqae
This is Thomas-Wigner rotation factor of Hilbert space matching~\cite{Chung:2019duq}, for which we now derive its rotation angle. 
\paragraph{Thomas-Wigner rotation} 
Let $u$, $v$, $w$ be 4-velocity vectors; each one is time-like, unit-normalized and future-pointing. 
Any two of them can be connected by a {\em minimal} boost: 
\begin{align}
u^\mu = G(u,v)^\m{}_\n v^\n \,.
\end{align} 
It is minimal in the sense that it acts non-trivially only on the 2-plane spanned by $u$ and $v$. 
This minimality fixes $G$ uniquely, and the explicit form is given by 
\begin{align}
G(u,v)^\m{}_\n = \delta^\m{}_\n - \frac{(u+v)^\m (u+v)_\n}{1+u \cdot v} + 2 u^\m v_\n \,.
\label{boost-explicit}
\end{align}
The inversion of $G$ exchanges the roles of $u$ and $v$:
\begin{align}
G(u,v)^{-1} = G(v,u) \,.
\end{align}
Now consider a closed loop of three minimal boosts, $G(u,v)G(v,w)G(w,u)$. Since it takes $u$ back to itself, the result should be a rotation on the 3-plane orthogonal to $u$. In a suitably chosen basis, the rotation would be represented by
\begin{align}
[G(u,v)G(v,w)G(w,u)]^\m{}_\n = 
\begin{pmatrix}
1 & 0 & 0 & 0 \\
0 & \cos\alpha & -\sin\alpha & 0 \\
0 & \sin\alpha & \cos\alpha & 0 \\
0 & 0 & 0 & 1 
\end{pmatrix} \,.
\end{align}
A manifestly Lorentz-invariant way to characterize the angle $\alpha$ is 
\begin{align}
\tr [G(u,v)G(v,w)G(w,u)] = 2 + 2 \cos\alpha\,.
\label{tr-cos-a}
\end{align}
Taking the trace explicitly using \eqref{boost-explicit}, we reproduce a well-known formula for the angle:
\begin{align}
2 + 2 \cos\alpha = 2 \frac{ (1+ u\cdot v+v\cdot w + w\cdot u)^2}{(1+u\cdot v)(1+v\cdot w)(1+w\cdot u)}\,.
\label{tr-cos-b}
\end{align}
We find it useful to rewrite \eqref{tr-cos-b} as 
\begin{align}
1 - \cos\alpha = \frac{- (\ve_{\m\n\r\s} u^\n v^\r \w^\s)^2}{(1+u\cdot v)(1+v\cdot w)(1+w\cdot u)} \,.
\label{alpha-great}
\end{align}
The $(-)$ sign on the RHS reflects the fact that the vector $\ve_{\m}(u,v,w) \equiv \ve_{\m\n\r\s} u^\n v^\r \w^\s$ is space-like when $u$, $v$, $w$ are time-like. Eq.~\eqref{alpha-great} clearly shows that the angle $\alpha$ vanishes when $u$, $v$, $w$ are linearly dependent. 

\paragraph{Scattering kinematics in the COM frame} 

Let us now specialize to the kinematics of the two body scattering \eqref{kin-scattering}. To compute the rotation angle $\alpha$ for particle $a$, we identify the velocity vectors to be
\begin{align}
u =\frac{p_{\rm in}}{m_a} =\frac{p_1}{m_a} \,,
\quad 
v = \frac{p_{\rm out}}{m_a}=\frac{p_2}{m_a} \,,
\quad 
w = \frac{p_a + p_b}{E_a+E_b} \,.
\label{velo-scattering}
\end{align}

We may insert \eqref{kin-scattering} and \eqref{velo-scattering} into \eqref{alpha-great}. 
For the denominator, we have 
\begin{align}
1 + u\cdot v = 2 \,,
\quad 
1 + u \cdot w = 1 + \frac{E_a}{m_a} =  1 + v \cdot w  \,.
\end{align}
For the numerator, we note that 
\begin{align}
\ve_\m(p_1,p_2,p_a + p_b) =  \ve_\m(p_a + q/2 ,p_a - q/2 ,p_a + p_b) = 
\ve_\m(p_a ,p_b , q) \,.
\end{align}
Combining all the ingredients, we obtain 
\begin{align}
2(1-\cos\alpha) = 4 \sin(\alpha/2)^2 = \frac{- [\ve_\m(p_a ,p_b, q)]^2}{m_a^2 E^2(m_a+E_a)^2} \,.
\end{align}
Let $f(x)$ be the inverse function of $2\sin(x/2)$. Clearly, $\tilde{f}(x^2) \equiv (f(x)-x)/x$ is an analytic function of $x^2$ with $\tilde{f}(0)=0$. 
Under the presumption of the HCL kinematics, since 
\begin{align}
(\ve_\m(p_a ,p_b , q))^2 \propto \left[p_a^2 p_b^2 - (p_a\cdot p_b)^2\right] q^2 \approx 0\,,
\end{align}
we may set $\tilde{f}(x^2) \approx 0$ and hence $f(x) \approx x$ in what follows. 

So far, we have worked out the magnitude of the angle $\alpha$ only. We should also find the orientation of the rotation plane. To put the incoming and out-going states on a nearly equal footing, we work in the COM frame. Then, the three 4-vectors $u_a$, $u_b$, $q$ together determine the rotation axis through the $\varepsilon$-tensor. For a spinor in 3d, the rotation is represented by
\begin{align}
U(\pm \hat{m},\alpha) = e^{\pm \frac{i}{2}\alpha (\hat{m}\cdot \vec{\sigma})} = e^{ \pm i \alpha (\hat{m}\cdot \vec{s})} \,.
\end{align}
We conclude that the rotation factor is
\begin{align}
U_{\rm rotation}^{(a)} = \exp\left[ - i \left(\frac{m_b}{r_a E}\right) \ve(q,u_a,u_b,a_a) \right] \,,
\quad 
r_a \equiv 1 + \frac{E_a }{m_a}  \,.
\label{phase-final}
\end{align}
We have fixed the sign in the exponent of \eqref{phase-final} by matching against our earlier 
work on the leading PN, all order in spin, computation \cite{Chung:2018kqs,Chung:2019duq}.

\subsection{Complete 1PM potential}
Equipped with the rotation factors $U_{\rm rotation}^{(a)} $ and $U_{\rm rotation}^{(b)}$, we simply dress the bare amplitude in eq.(\ref{spin-amp-BH}) for black holes as 
\begin{align}
M_{\rm dressed}^{\rm (BH)} = -\frac{4\pi G}{q^2} \frac{m_a^2 m_b^2}{E_aE_b} 
\cosh\left( 2\theta  +  i \frac{\ve(q,u_a,u_b,a_0)}{\sinh\theta} \right) U^{(a)} U^{(b)} \,,
\label{BH-dressed}
\end{align}
where we suppress the subscript on $U$. The expression combines the amplitude eq.\eqc{spin-amp-BH} with additional rotation factors $U^{(a)} U^{(b)}$ originating from how polarization tensors are defined.
Setting $q= (0, \vec{q})$ and taking the Fourier transform with $e^{i\vec{q}\cdot \vec{r}}$, we obtain the potential. Since an exponentiated gradient generates a finite translation, 
we can explicitly write the potential as 
\begin{align}
V_{\rm 1PM}^{\rm (BH)} = - \frac{G m_a^2m_b^2}{2E_a E_b} 
\sum_{s=\pm 1} 
e^{2s\theta}
\left| \vec{r} + s\frac{E (\vec{p}\times \vec{a}_0)}{m_a m_b \sinh\theta} - \frac{\vec{p}\times \vec{a}_a}{m_a r_a} - \frac{\vec{p}\times \vec{a}_b}{m_b r_b} 
\right|^{-1} \,. 
\label{V-BH-final}
\end{align}
For general compact spinning bodies with non-minimal Wilson coefficients, 
we dress the general form of the amplitude \eqref{spin-amp-bare} 
with the rotation factors  
to reach the master formula:
\begin{align}
\begin{split}
V_{\rm 1PM}^{\rm (general)} &= - \frac{4\pi G m_a^2 m_b^2}{E_a E_b} \int \frac{d^3 \vec{q}}{(2\pi)^3} e^{i\vec{q}\cdot \vec{r}}
\left[\frac{1}{2} \sum_{s=\pm 1} e^{2s\theta} W_a(s\tau_a) W_b(s\tau_b)\right] U^{(a)} U^{(b)} \,.
\label{V-1PM-master}
\end{split}
\end{align}
We can still perform the Fourier transform, but the result is not as simple as \eqref{V-BH-final}.

In the next section, we expand our master formula in (\ref{V-1PM-master}) up to quartic order in spin and obtain an explicit and exact expression in $\vec{p}$ at each spin order.  As confirmed already in \cite{Chung:2018kqs,Chung:2019duq}, the leading order terms in the $\vec{p}$ expansion are free from ambiguities and are easily shown to agree with well established results in the literature. In section~\ref{sec:checks}, as further checks, we expand our expressions to higher orders in $\vec{p}$ and compare with NLO and NNLO PN results available in the literature.

\section{1PM potential at each spin order} \label{sec:spin-order}
In this section, we present an explicit form of the 1PM for potential at each fixed order of spin. The exact result and the LO term will be presented here, while matching at NLO and beyond will be the focus of the next section. To demonstrate the almost complete factorization of the spin-dependence and the momentum dependence, we organize the results using the following notations. 
In writing down the spin(a)$^m$-spin(b)$^n$ term $V_{S_a^m S_b^n}$ of the potential eq.\eqc{V-1PM-master}
, we write
\begin{align}
\begin{split}
V_{S_a^m S_b^n} &= \left(\frac{G m_a m_b }{r^{m+n+1}} \right) \left[ F_{(m,n)}(\vec{a}_a,\vec{a}_b,\hat{n}) \right] X_{(m,n)}(\vec{p}^2) 
\qquad (m+n\;\; \mbox{even}) \,,
\\
V_{S_a^m S_b^n} &= \left(\frac{G m_a m_b}{r^{m+n+1}} \right) \left[ \vec{p}\cdot \vec{F}_{(m,n)}(\vec{a}_a,\vec{a}_b,\hat{n}) \right]  X_{(m,n)}(\vec{p}^2) 
\quad (m+n\;\; \mbox{odd}) \,.
\end{split}
\label{def:FvF}
\end{align} 
Explicitly, the spin-dependent factors, $F_{(m,n)}$ and $\vec{F}_{(m,n)}$, are defined by
\begin{align}
\begin{split}
F_{(m,n)} &= r^{m+n} (\vec{a}_a \cdot \nabla)^m (\vec{a}_b \cdot \nabla)^n \left(\frac{1}{r} \right) \,,
\\
\vec{F}_{(m,n)} &= \left\{
\begin{aligned}
\frac{r^{m+n}}{m_a} (\vec{a}_a \times \nabla)  (\vec{a}_a \cdot \nabla)^{m-1} (\vec{a}_b \cdot \nabla)^n \left(\frac{1}{r} \right) && \qquad (m\;\; \mbox{odd}) \,,
\\
\frac{r^{m+n}}{m_b} (\vec{a}_b \times \nabla)  (\vec{a}_a \cdot \nabla)^{m} (\vec{a}_b \cdot \nabla)^{n-1} \left(\frac{1}{r} \right) && \qquad (n\;\; \mbox{odd}) \,.
\end{aligned} \right.
\end{split}
\end{align}
By construction, $F_{(m,n)}$ and $\vec{F}_{(m,n)}$ are homogeneous polynomials of $\vec{a}_a$ and $\vec{a}_b$ of degree $m$ and $n$, respectively. We pulled out an overall factor of masses so as to make $X_{(m,n)}$ dimensionless. 
When we expand the potential in $\vec{p}$, we will use the notation 
\begin{align}
X_{(m,n)} = X_{(m,n)}^{\rm LO} + X_{(m,n)}^{\rm NLO} + \cdots + X_{(m,n)}^{{\rm N}^k{\rm LO} } + \cdots \,,
\end{align}
where $X_{(m,n)}^{{\rm N}^k{\rm LO}}$ is proportional to $(\vec{p}^{2})^k$.

Regardless of the order of expansion in spin or momentum, 
there are two notable differences between our result and those in the literature. 
First, ours results doesn't carry any $(\hat{n}\cdot \vec{p})$ term. 
(Here $\hat{n} = \vec{r}/{r}$ is the unit directional vector between the two bodies.)
In other words, the so-called ``isotropic gauge" is forced upon us 
by the amplitude approach; see \cite{Bern:2019crd} for a related comment.
Second, ours results doesn't carry any $(\vec{a} \cdot \vec{p})$ term either, except 
through a very specific $(\vec{p}\cdot \vec{F})$ structure in \eqref{def:FvF}. 
This is to be contrasted with a typical PN computation which often produces 
a linear combination, 
\begin{align}
\vec{a}_a^2 f_1 (\vec{p}^2) + (\vec{a}_a \cdot \vec{p})^2 f_2(\vec{p}^2)  \,,
\end{align}
with no obvious correlation between the two functions $f_1$ and $f_2$. 

A minor technical remark. To reduce clutter in equations, we introduce a few more short-hand notations such as
$s_\theta = \sinh\theta$, $c_\theta = \cosh\theta$, $c_{2\theta} = \cosh(2\theta)$. 

\subsection{Linear in spin}
Since the Wilson coefficients $C_0=C_1=1$ are universal, at linear order in spin the potential is universal and we may simply work with the black holes. First, from $\mathcal{M}_{\rm bare, BH}$ in eq.(\ref{spin-amp-BH}) the spin-linear term is
\begin{align}
\begin{split}
\cosh\left( 2\theta  +  i \frac{\ve(q,u_a,u_b,a_0)}{\sinh\theta} \right) 
&= \cosh(2\theta) + i  \frac{\ve(q,u_a,u_b,a_0)}{\sinh\theta} \sinh2\theta + \mathcal{O}(a_0^2)
\\
&\approx  \cosh(2\theta) + 2 i (\cosh\theta) \ve(q,u_a,u_b,a_0) \,.
\end{split}
\end{align}
Using the identity,  
\begin{align}
\ve(q,u_a,u_b,a) = \left( \frac{E}{m_a m_b} \right) \vec{p} \cdot (\vec{a} \times \vec{q}) \,,
\end{align}
we find the contribution from $\mathcal{M}_{\rm bare}$ to the spin-linear potential can be written as: 
\begin{align}
V_{\rm bare} = - \frac{2m_a m_b E }{E_a E_b} (\cosh\theta) 
\left[ \vec{p} \cdot (\vec{a}_0 \times \nabla) \right] \left( \frac{G}{r} \right) \,.  
\end{align}
Now from the rotation factor $U^{(a)}$, we find 
\begin{align}
\begin{split}
V_{\rm rotation}
&= \frac{m_a^2m_b^2}{E_a E_b} \frac{1}{m_ar_a } \cosh(2\theta) 
\left[ \vec{p} \cdot (\vec{a}_a \times \nabla) \right] \left( \frac{G}{r} \right) 
\\
&= \frac{m_a m_b^2 }{E_a E_b r_a } \cosh(2\theta) 
\left[ \vec{p} \cdot (\vec{a}_a \times \nabla) \right] \left( \frac{G}{r} \right) 
\,.
\end{split}
\end{align}
Collecting all the terms we obtain 
\begin{align}
V_{S_a^1 S_b^0} = - \frac{m_a m_b E}{E_a E_b} \left(2 c_\theta  - \frac{m_bc_{2\theta}}{Er_a } \right) \left[ \vec{p} \cdot (\vec{a}_a \times \nabla) \right] \left( \frac{G}{r} \right) \,,
\end{align}
or, equivalently, 
\begin{align}
V_{S_a^1 S_b^0} = \left( \frac{G}{r^2} \right)  \frac{m_a m_b E}{E_a E_b} \left(2 c_\theta - \frac{m_bc_{2\theta}}{Er_a } \right) 
[ \vec{p}  \cdot ( \vec{a}_a \times \hat{n} ) ]
\,.
\label{SO-exact}
\end{align}
In the notation of \eqref{def:FvF}, we have
\begin{align}
\vec{F}_{(1,0)} =  \vec{a}_a \times \hat{n} \,,
\quad 
X_{(1,0)} = \frac{m_a E}{E_a E_b} \left(2 c_\theta  - \frac{m_bc_{2\theta}}{Er_a } \right)  \,.
\end{align}
This is the exact linear in spin potential at 1PM. 

\paragraph{LO}
The leading order term in $\vec{p}^2$ can be extracted and given by:
\begin{align}
V_{S_a^1 S_b^0}^{\rm LO} = 
\left( \frac{G}{r^2} \right) [ \vec{p}  \cdot ( \vec{a}_a \times \hat{n} ) ] \left( \frac{4 m_a + 3 m_b}{2}\right)
\,.
\end{align}

\subsection{Quadratic in spin}
\paragraph{Spin-spin couplings}

This term also only utilizes $C_0=C_1=1$ only and thus are universal as well. From the 
bare amplitude $\mathcal{M}_{\rm bare}$
\begin{align}
(V_{\rm bare})_{S_a^1S_b^1} = - c_{2\theta}  \left(\frac{G}{r^3}\right) \frac{m_a^2m_b^2}{E_a E_b} 
 \left[\vec{a}_a\cdot\vec{a}_b  - 3 (\vec{a}_a\cdot \hat{n}) (\vec{a}_b \cdot \hat{n})\right] \,.
\end{align}
Adding up the other two contributions, we obtain 
\begin{align}
\begin{split}
V_{S_a^1 S_b^1} &=  \left(\frac{Gm_a m_b }{r^3}\right) F_{(1,1)} X_{(1,1)}  \,, 
 \quad
 F_{(1,1)} =  - \left[\vec{a}_a\cdot\vec{a}_b  - 3 (\vec{a}_a\cdot \hat{n}) (\vec{a}_b \cdot \hat{n})\right] \,,
 \\
 X_{(1,1)} &= \frac{m_a m_b }{E_a E_b} \left[ c_{2\theta} -\frac{2s_\theta^2 c_{\theta}}{E}
 \left(\frac{m_b}{r_a} +\frac{m_a}{r_b}\right) 
 + \frac{m_a m_b s_\theta^2 c_{2\theta}}{E^2r_a r_b} \right]
\,.
\end{split}
\label{SS-ab}
\end{align}

\paragraph{Spin-squared} 

For the spin-squared piece, one has $C_2$ contribution from $\mathcal{M}_{\rm bare}$, as well as $C_1$ from $\mathcal{M}_{\rm bare}$ times linear expansion of $U$ and the quadratic in spin expansion of $U$. The latter two are once again universal. Adding up all three contributions, we obtain
\begin{align}
\begin{split}
V_{S_a^2 S_b^0} &=  \left(\frac{Gm_a m_b }{r^3}\right) F_{(2,0)} X_{(2,0)}  \,, 
\quad
F_{(2,0)} = - \left[\vec{a}_a^2  - 3 (\vec{a}_a\cdot \hat{n})^2  \right]  \,,
\\
X_{(2,0)} &=  \frac{m_a^2m_b^2}{2 E_a E_b} \left[ C_2^{(a)} c_{2\theta} 
- \frac{4m_b s_\theta^2  c_{\theta}}{Er_a } 
+ \frac{m_b^2 s_\theta^2 c_{2\theta}}{E^2r_a ^2}  \right]
\,.
\end{split}
\label{SS-aa}
\end{align}

\paragraph{LO} To the leading order we have:
\begin{align}
\begin{split}
V^{\rm LO}_{S_a^1 S_b^1} &= - \left(\frac{G m_a m_b}{r^3}\right)  \left[\vec{a}_a\cdot\vec{a}_b  - 3 (\vec{a}_a\cdot \hat{n}) (\vec{a}_b \cdot \hat{n})\right] \,,
\\
V^{\rm LO}_{S_a^2 S_b^0} &= -  \frac{1}{2} C_2^{(a)} \left(\frac{G m_a m_b}{r^3}\right)  \left[\vec{a}_a^2  - 3 (\vec{a}_a\cdot \hat{n})^2  \right] \,.
\end{split}
\end{align}

\subsection{Cubic in spin}
Continuing with the same method, we obtain the formulae for the cubic-in-spin terms. 
For the spin(a)$^3$ term, we have 
\begin{align}
\begin{split}
V_{S_a^3 S_b^0} &= \left(\frac{Gm_b}{r^4}\right) [\vec{p}\cdot \vec{F}_{(3,0)}] X_{(3,0)}  \,,
\\
\vec{F}_{(3,0)} &= 3 (\vec{a}_a \times \hat{n}) \left[\vec{a}_a^2  - 5 (\vec{a}_a\cdot \hat{n})^2  \right] \,,
\\
X_{(3,0)}&= \frac{m_a E}{E_a E_b} \left[
\frac{1}{3} C_3^{(a)}  c_{\theta} - C_2^{(a)} \frac{m_b c_{2\theta}}{2Er_a } 
+  \frac{m_b^2 \sinh ^2\theta  \cosh \theta }{E^2r_a ^2 }
-\frac{m_b^3 \sinh ^2\theta  \cosh (2\theta) }{6E^3r_a^3 } \right]  \,.
\end{split}
\label{SSS-aaa}
\end{align}
For the mixed spin(a)$^2$-spin(b)$^1$ term, we have
\begin{align}
\begin{split}
V_{S_a^2 S_b^1} &=  \left(\frac{Gm_b}{r^4}\right) 
[\vec{p}\cdot \vec{F}_{(2,1)} ] X_{(2,1)} \,,
\\
\vec{F}_{(2,1)} &= 3\left\{ (\vec{a}_b \times \hat{n} ) \left[\vec{a}_a^2  - 5 (\vec{a}_a\cdot \hat{n})^2 \right]  -2 (\vec{a}_a\cdot \hat{n}) (\vec{a}_a\times \vec{a}_b)\right\} \,,
\\
X_{(2,1)}&= \frac{m_a  E}{E_a E_b} \left[ C_2^{(a)} \cosh \theta - \frac{c_{2\theta}}{2E}\left(\frac{2m_b}{r_a }+ C_2^{(a)} \frac{m_a}{r_b} \right) 
\right.
\\
&\qquad \qquad\qquad
\left.
+ \frac{m_b s^2_\theta   c_{\theta} }{E^2r_a } 
\left( \frac{m_b  }{r_a } +\frac{2 m_a }{r_b}\right)
-\frac{m_a m_b^2 s^2_\theta c_{2\theta} }{2 E^3r_a ^2 r_b} \right]\,.
\end{split}
\label{SSS-aab}
\end{align}

\paragraph{LO} To the leading order, we find 
\begin{align}
\begin{split}
V_{S_a^3 S_b^0}^{\rm LO} &= 
\left(\frac{G}{r^4}\right) 
\vec{p}  \cdot ( \vec{a}_a \times \hat{n} )   \left[\vec{a}_a^2  - 5 (\vec{a}_a\cdot \hat{n})^2  \right] 
\left( C_3^{(a)} (m_a+m_b) - \frac{3}{4} C_2^{(a)} m_b \right) \,,
\\
V_{S_a^2 S_b^1}^{\rm LO} &=\left(\frac{G}{r^4}\right) \vec{p}\cdot \left\{ ( \vec{a}_b \times \hat{n} ) \left[\vec{a}_a^2  - 5 (\vec{a}_a\cdot \hat{n})^2 \right]  -2 (\vec{a}_a\cdot \hat{n}) (\vec{a}_a\times \vec{a}_b)
\right\}
\\
&\qquad \times 
 \left(  \frac{3}{4} C_2^{(a)}(3m_a + 4m_b)- \frac{3}{2} m_b\right)  \,,
\end{split}
\end{align}
in perfect agreement with the corresponding terms in eq.(3.10) of \cite{Levi:2014gsa}.

\subsection{Quartic in spin} 
We continue to quartic in spin. This is an interesting threshold for black holes from an on-shell perspective, since fundamental massive particles are only known up to spin-2. It can be shown that beyond spin-2, isolated spinning particle no-longer exists and must either be a bound state or part of an infinite tower of massive states~\cite{Arkani-Hamed:2017jhn, Afkhami-Jeddi:2018apj}.  As a consequence, the gravitational Compton amplitude is no longer unique beyond spin-2. However, this ambiguity only arrises at 2PM.

The spin(a)-quartic term is 
\begin{align}
\begin{split}
V_{S_a^4 S_b^0} 
&= \left(\frac{Gm_a m_b}{r^5}\right) F_{(4,0)} X_{(4,0)} \,,
\\
F_{(4,0)} &= 3 \left\{ 3\vec{a}_a^4 -30 \vec{a}_a^2 (\vec{a}_a\cdot \hat{n})^2 + 35(\vec{a}_a\cdot \hat{n})^4 \right\}  \,,
\\
X_{(4,0)}&= -\frac{m_a m_b}{24 E_a E_b} \left[ c_{2\theta}
\left(
C_4^{(a)} + 6 C_2^{(a)} \frac{ m_b^2 s^2_\theta  }{r_a ^2 E^2} + \frac{m_b^4 s^4_\theta }{r_a ^4 E^4} \right)
-\frac{8 m_b s^2_\theta c_\theta}{r_a  E}
\left(
C_3^{(a)} + \frac{m_b^2 s^2_\theta}{r_a ^2 E^2} 
\right)
\right]
\,.
\end{split}
\label{S4-aaaa}
\end{align}
The cubic-linear term is
\begin{align}
\begin{split}
V_{S_a^3 S_b^1} 
&= \left(\frac{Gm_a m_b}{r^5}\right)  F_{(3,1)} X_{(3,1)}  \,,
\\
F_{(3,1)} &= 3\left\{ 3\vec{a}_a^2(\vec{a}_a\cdot \vec{a}_b) -15 (\vec{a}_a\cdot\vec{a}_b) (\vec{a}_a\cdot \hat{n})^2 \right.
\left. -15 \vec{a}_a^2(\vec{a}_a\cdot \hat{n})(\vec{a}_b\cdot \hat{n}) + 35(\vec{a}_a\cdot \hat{n})^3(\vec{a}_b\cdot \hat{n}) \right\}\,,
\\
X_{(3,1)}&= -\frac{m_a m_b}{6E_aE_b} \left[ c_{2 \theta} \left\{ C_3^{(a)} 
+\frac{3 m_b s^2_\theta }{r_a  E^2} \left(C_2^{(a)} \frac{m_a}{r_b}+\frac{m_b}{r_a}\right)
+\frac{m_a m_b^3 s^4_\theta}{r_a ^3 r_b E^4}
\right\} \right.
\\
&\qquad \qquad \qquad \qquad 
\left. -
\frac{2s^2_\theta c_\theta}{E}
\left\{ C_3^{(a)} \frac{ m_a}{r_b} + 3 C_2^{(a)} \frac{m_b}{r_a}+\frac{m_b^2 s^2_\theta }{r_a ^2 E^2} \left(\frac{3 m_a}{r_b}+\frac{m_b}{r_a}\right)\right\} \right]
 \,.
\end{split}
\label{S4-aaab}
\end{align}
The quadratic-quadratic term is the first place where non-trivial Wilson coefficients 
from both spinning bodies contribute together. 
\begin{align}
\begin{split}
V_{S_a^2 S_b^2} 
&=  \left(\frac{Gm_am_b}{r^5}\right) F_{(2,2)} X_{(2,2)}
 \,,
\\
F_{(2,2)} &= 3 \left\{ \vec{a}_a^2 \vec{a}_b^2 + 2(\vec{a}_a\cdot \vec{a}_b)^2 
-5 \vec{a}_a^2  (\vec{a}_b\cdot \hat{n})^2 -5 \vec{a}_b^2  (\vec{a}_a\cdot \hat{n})^2 \right.
\\
&\qquad\qquad\qquad\qquad 
\left. -20 (\vec{a}_a\cdot \vec{a}_b) (\vec{a}_a\cdot \hat{n})(\vec{a}_b\cdot \hat{n}) 
+ 35(\vec{a}_a\cdot \hat{n})^2(\vec{a}_b\cdot \hat{n})^2 \right\} \,,
\\
X_{(2,2)}&= -\frac{m_am_b}{4E_aE_b} \left[ c_{2\theta}
\left\{
\left(C_2^{(a)}+\frac{m_b^2 s^2_\theta}{r_a ^2 E^2}\right) 
\left(C_2^{(b)}+\frac{m_a^2 s^2_\theta}{r_b^2 E^2}\right) 
+ \frac{4 m_a m_b s^2_\theta }{r_a  r_b E^2}
\right\} \right.
\\
&\qquad \qquad \qquad
\left.
- \frac{4s^2_\theta c_\theta}{E} 
\left\{ C_2^{(a)} \frac{m_a}{r_b}+C_2^{(b)} \frac{m_b}{r_a}
+ \frac{m_a m_b s^2_\theta}{r_a  r_b E^2} \left( \frac{m_a}{r_b} + \frac{m_b}{r_a}\right) \right\} \right]
 \,.
\end{split}
\label{S4-aabb}
\end{align}

\paragraph{LO} 
To leading order, we find
\begin{equation}
\begin{split}
V_{S_a^4,S_b^0} &= -\frac{G m_a m_b}{8r^5} C_4^{(a)}\left\{ 3\vec{a}_a^4 -30 \vec{a}_a^2 (\vec{a}_a\cdot \hat{n})^2 + 35(\vec{a}_a\cdot \hat{n})^4 \right\}\\
V_{S_a^3,S_b^1} &= -\frac{G m_a m_b}{2r^5} C_3^{(a)}\left\{ 3\vec{a}_a^2(\vec{a}_a\cdot \vec{a}_b) -15 (\vec{a}_a\cdot\vec{a}_b) (\vec{a}_a\cdot \hat{n})^2 \right.\\
&\qquad\qquad\qquad\qquad\qquad\qquad\quad
\left. -15 \vec{a}_a^2(\vec{a}_a\cdot \hat{n})(\vec{a}_b\cdot \hat{n}) + 35(\vec{a}_a\cdot \hat{n})^3(\vec{a}_b\cdot \hat{n}) \right\}\\
V_{S_a^2,S_b^2} &= -\frac{3G m_a m_b}{4r^5} C_2^{(a)}\left\{ \vec{a}_a^2 \vec{a}_b^2 + 2(\vec{a}_a\cdot \vec{a}_b)^2 
-5 \vec{a}_a^2  (\vec{a}_b\cdot \hat{n})^2 -5 \vec{a}_b^2  (\vec{a}_a\cdot \hat{n})^2 \right.
\\
&\qquad\qquad\qquad\qquad \qquad
\left. -20 (\vec{a}_a\cdot \vec{a}_b) (\vec{a}_a\cdot \hat{n})(\vec{a}_b\cdot \hat{n}) 
+ 35(\vec{a}_a\cdot \hat{n})^2(\vec{a}_b\cdot \hat{n})^2 \right\}
\end{split}
\end{equation}
which is in perfect agreement to eq.(4.4) of \cite{Levi:2014gsa}.

\section{Reproducing 1PM part of PN expansion} \label{sec:checks}

In the previous section, we derived the potential at each spin order that is exact in $\vec{p}$. It is almost trivial to expand the expressions in powers of $\vec{p}^2$. 
Each term in the $\vec{p}^2$ expansion can be compared with the 1PM part of the 
PN computation available in the literature. In this section, we make the comparison explicitly for all spin and momentum orders where the data are available. 

The precise form of the subleading terms in $\vec{p}^2$ depend on the choice of the phase space coordinates $(\vec{r},\vec{p})$. This ``coordinate gauge" ambiguity originates from the general covariance of general relativity. Any two different gauge choices are related to each other by a canonical transformation. We denote by $g$ the generator 
of the canonical transformation, 
\begin{align}
\Delta_\epsilon H = \epsilon \{ H, g \} \,,
\label{cano-generator-epsilon}
\end{align}
where $\epsilon$ is an infinitesimal parameter. In the PN expansion, both $G$ and $1/m$ can be treated as if they were infinitesimal, so we will use a variant of \eqref{cano-generator-epsilon} without explicitly mentioning the infinitesimal parameter $\epsilon$. 

At NLO in the PN expansion, the only relevant term in $H$ on the right-hand side of \eqref{cano-generator-epsilon} is the Newtonian term $H_{\rm N}$. Since we are 
comparing terms at 1PM only, only the kinetic term of $H_{\rm N}$ contribute. 
\begin{align}
(\Delta H^{\rm NLO})_{\rm 1PM} = \{ H_{\rm N} , g^{\rm NLO} \} = \left\{ \left(\frac{1}{2m_a} + \frac{1}{2m_b} \right) \vec{p}^2 , g^{\rm NLO} \right\} 
+ \mathcal{O}(G^2) \,. 
\label{cano-NLO}
\end{align}
The transformation receives two contributions at NNLO. 
\begin{align}
\begin{split}
(\Delta H^{\rm NNLO})_{\rm 1PM} &= \{ H_{\rm N} , g^{\rm NNLO} \} + \{ H_{\rm 1PN} , g^{\rm NLO} \} + \mathcal{O}(G^2)\,,
\\
H_{\rm 1PN} &= - \left( \frac{1}{8m_a^3} +\frac{1}{8m_b^3} \right)\vec{p}^4 + \mathcal{O}(G) \,.
\end{split}
\label{cano-NNLO}
\end{align}
All canonical transformations to be performed below are based on the elementary Poisson 
algebra: 
$\{ x^i , p_j \} = \delta^i{}_j$. 
The following formula will be used multiple times:
\begin{align}
\begin{split}
&\left\{ \frac{\vec{p}^2}{2}  , \left(\frac{1}{r}\right)^k (\hat{n}\cdot \vec{p})^\ell [\hat{n}\cdot (\vec{p}\times \vec{a})]^m \right\} 
\\
&= 
\left(\frac{1}{r}\right)^{k+1} (\hat{n}\cdot \vec{p})^{\ell -1} [\hat{n}\cdot (\vec{p}\times \vec{a})]^m \left( (k+\ell+m)(\hat{n} \cdot \vec{p})^2 - \ell  \vec{p}^2 \right) 
\,.
\end{split}
\label{cano-master}
\end{align}

\subsection{Linear in spin (up to NNNLO)} 

As explained earlier, our notation for the 1PM and arbitrary PN expansion is 
\begin{align}
V_{S_a^1 S_b^0} = 
\left( \frac{G m_b}{r^2} \right) [ \vec{p}  \cdot ( \vec{a}_a \times \hat{n} ) ] 
\left( X_{(1,0)}^{\rm LO} + X_{(1,0)}^{\rm NLO} + X_{(1,0)}^{\rm NNLO} + X_{(1,0)}^{\rm NNNLO} + \cdots \right)
\,.
\nonumber
\end{align}

\paragraph{NLO and its canonical transformations}

Expanding our formula \eqref{SO-exact}, we find 
\begin{align}
X_{(1,0)}^{\rm NLO} =  
\left( \frac{ 18 m_a^2  + 8 m_a m_b -5 m_b^2 } {8m_a^2 m_b^2} \right)\vec{p}^2 \,.
\label{NLO-SO-iso}
\end{align}
This NLO spin-orbit coupling was computed in the ADM framework in \cite{Damour:2007nc,Steinhoff:2008zr}, 
in the EFT framework in \cite{Levi:2014sba}, and in an amplitude-based approach in \cite{Vaidya:2014kza}. 
The last reference employs the isotropic gauge and the result looks identical to ours. 
It also explains how to use a canonical transformation to check agreement with \cite{Damour:2007nc,Steinhoff:2008zr}. 

Consider a family of Hamiltonians: 
\begin{align}
\begin{split}
H_{\rm SO}^{\rm NLO} &= \left( \frac{G}{r^2} \right) \frac{\hat{n}\cdot(\vec{p}\times \vec{a}_a)}{ 8m_a}
\left[  h_1 \vec{p}^2  + h_2 (\hat{n}\cdot\vec{p})^2  
\right]
\,,
\\
h_k &= h_{k,+} \zeta + h_{k,0} + h_{k,-} \zeta^{-1} \,, \quad (\zeta \equiv m_b/m_a)
\,.
\end{split}
\label{NLO-SO-family}
\end{align}
In this notation, our result \eqref{NLO-SO-iso} amounts to 
\begin{align}
h_1 =  -5 \zeta + 8 + 18 \zeta^{-1} \,,
\quad 
h_2 = 0 \,.
\label{h-so-ours}
\end{align}
Not all parameters are physically meaningful, because some combinations can be altered by 
canonical transformations of the type shown in \eqref{cano-NLO}.
Taking hints from \cite{Vaidya:2014kza}, we take the following ansatz for the generator of the transformation: 
\begin{align}
g_{\rm SO}^{\rm NLO} &= \frac{g_1}{8(1+\zeta^{-1})} \left( \frac{G}{r}\right) [\hat{n}\cdot (\vec{p}\times \vec{a}_a  )]  
(\hat{n}\cdot \vec{p}) \,,
\quad 
g_1 = g_{1,+} \zeta + g_{1,0} + g_{1,-} \zeta^{-1}\,.
\label{gen-SO-NLO}
\end{align}
The factor $1/(1+\zeta^{-1})$ in the generator is inserted to cancel the similar factor in \eqref{cano-NLO}.

Recalling the formula \eqref{cano-master} and setting $k=\ell=m=1$, we can express the changes $\Delta h_k$ in terms of $g_k$:
\begin{align}
\begin{split}
\Delta h_1 = - g_1 \,,
\quad 
\Delta h_2 = 3 g_1 \,.
\end{split}
\label{dh-vs-g}
\end{align} 

Several papers report the NLO spin-orbit potential. For example, eq.(6.22) of \cite{Levi:2015msa}, after being simplified in the COM frame, gives
\begin{align}
h_1 =  -5 \zeta + 8 \zeta^{-1} \,,
\quad 
h_2 =  24 + 30 \zeta^{-1}  \,.
\label{h-so-LS}
\end{align}
Taking the difference, $\Delta h_k = h_k^{\rm old} - h_k^{\rm new}$,  between \eqref{h-so-ours} and \eqref{h-so-LS}, 
we find
\begin{align}
\Delta h_1 = -8 - 10 \zeta^{-1}\,,
\quad 
\Delta h_2 = 24 + 30 \zeta^{-1} \,.
\end{align}
This is compatible with \eqref{dh-vs-g} if we set $g_1 = 8 + 10 \zeta^{-1}$. Thus we have shown that \eqref{NLO-SO-iso} is equivalent to the corresponding term in \cite{Levi:2015msa}.

\paragraph{NNLO its canonical transformations}

\begin{align}
X_{(1,0)}^{\rm NNLO} = 
\left( \frac{- 15 m_a^4 - 15m_a^2 m_b^2  -12 m_a m_b^3 + 7m_b^4}{16m_a^4 m_b^4} \right) \vec{p}^4 \,.
\label{NNLO-SO}
\end{align}
The same term in the Hamiltonian formulation was computed in the ADM framework in \cite{Hartung:2011te,Hartung:2013dza} and in the EFT framework in \cite{Levi:2015uxa}.

Once again, consider the following ansatz for the Hamiltonian: 
\begin{align}
H_{\rm SO}^{\rm NNLO} &= \left( \frac{G}{r^2} \right) \frac{\hat{n}\cdot(\vec{p}\times \vec{a}_a)}{ 16 m_a^2 m_b}
\left[  h_3 \vec{p}^4  + h_4 \vec{p}^2 (\hat{n}\cdot\vec{p})^2 + h_5 (\hat{n}\cdot\vec{p})^4
\right]
\,.
\label{NLO-SO-family2}
\end{align}
In this notation, our result \eqref{NNLO-SO} correspond to 
\begin{align}
h_3 =  7 \zeta^2 - 12 \zeta -15 -15 \zeta^{-2} \,,
\quad 
h_4 = h_5 = 0 \,.
\label{h-so-ours2}
\end{align}
Eq.(4.11) of \cite{Levi:2015uxa}, sharing the same convention as \cite{Levi:2015msa}, is translated to our notation as 
\begin{align}
h_3 = 7\zeta^2 - 4 \zeta -24 -20 \zeta -12\zeta^{-2} \,,
\quad 
h_4 = -8\zeta -3 + 8\zeta^{-1} \,,
\quad 
h_5 = 60 + 60\zeta^{-1} -15\zeta^{-2}\,.
\end{align}
The difference between the two results is then 
\begin{align}
\Delta h_3 = 8 \zeta - 9 -20 \zeta + 3 \zeta^{-2} \,,
\quad 
\Delta h_4 = - 3(8\zeta +3 - 8\zeta^{-1}) \,,
\quad 
\Delta h_5 = 15( 4  + 4\zeta^{-1} -\zeta^{-2})\,.
\label{NNLO-data}
\end{align}

Our ansatz for the NNLO generating function is
\begin{align}
g_{\rm SO}^{\rm NNLO} &= \frac{1}{16(1+\zeta^{-1})} \left( \frac{G}{r}\right) 
\frac{[\hat{n}\cdot (\vec{p}\times \vec{a}_a  )]}{m_a m_b} 
\left[ g_2 \vec{p}^2 (\hat{n}\cdot \vec{p}) + g_3 (\hat{n}\cdot \vec{p})^3 \right] \,.
\label{gen-SO-NNLO}
\end{align}
Using \eqref{cano-NNLO} and \eqref{cano-master}, we can easily relate the 
coefficients, 
\begin{align}
\Delta h_3 =  (\zeta-1+\zeta^{-1}) g_1 - g_2\,,
\quad 
\Delta h_4 =  3 [ -(\zeta-1+\zeta^{-1}) g_1 +  g_2 - g_3 ] \,,
\quad 
\Delta h_5 =  5g_3 \,.
\label{NNLO-relation}
\end{align}
The value of $g_1$ was already fixed at the NLO order. The difference \eqref{NNLO-data} 
matches the relation \eqref{NNLO-relation} if we set
\begin{align}
g_2 = (1+\zeta^{-1})(11+7\zeta^{-1}) \,,
\quad 
g_3 = 3(4+4\zeta^{-1} - \zeta^{-2}) \,.
\end{align}

\paragraph{NNNLO} 

To the best of our knowledge, the NNNLO spin-orbit coupling has not been computed yet. We simply present the result.
\begin{align}
X_{(1,0)}^{{\rm N}^3{\rm LO}} = 
\left( \frac{84 m_a^6+50 m_a^4 m_b^2+84 m_a^2 m_b^4+80 m_a m_b^5-45 m_b^6}{128 m_a^6 m_b^6} \right) \vec{p}^6 \,.
\label{N3LO-SO}
\end{align}
%

\subsection{Quadratic in spin (up to NNLO)}

Expanding the exact results \eqref{SS-ab} and \eqref{SS-aa} in $\vec{p}^2$, 
we obtain sub-leading corrections. We write down our results explicitly 
up to NNLO and compare them with previous PN computations. 

The NLO spin-spin Hamiltonian was computed in the ADM framework in \cite{Steinhoff:2007mb,Hartung:2011ea} 
and in the EFT framework in \cite{Levi:2011eq,Levi:2014sba}. The equivalence 
between the two approaches was established in \cite{Levi:2014sba}.
The NLO spin-squared coupling was computed in \cite{Porto:2008jj,Steinhoff:2008ji,Hergt:2008jn,Hergt:2010pa,Levi:2015msa}. 
The NNLO spin-squred couplings were computed in
\cite{Levi:2015ixa, Levi:2016ofk}. The equivalence among different approaches 
were established in later references. 

\paragraph{NLO} 
The NLO spin-spin term in our framework is
\begin{align}
X^{\rm NLO}_{(1,1)} = 
\left( \frac{ 2 m_a^2  + 9 m_a m_b + 2 m_b^2  }{4m_a^2 m_b^2} \right) \vec{p}^2 \,.
\label{NLO-SSb-iso}
\end{align}
It can be compared with eq.~(6.32) of \cite{Levi:2015msa}. 
Even after reducing to the COM frame, the result of \cite{Levi:2015msa} appears 
to carry many non-vanishing coefficients. 
It is not clear how many of them are gauge invariant. 
According to our result, only two of them are invariant once we take into account 
the exchange symmetry, $m_a \leftrightarrow m_b$. 

The NLO spin-squared term in our framework is
\begin{align}
X_{(2,0)}^{\rm NLO} &= 
\left( \frac{C_2^{(a)} (6m_a^2 +16 m_a m_b + 6 m_b^2) - (8m_a + 7 m_b) m_b }{8m_a^2 m_b^2}
\right) \vec{p}^2 
\,.
\label{NLO-SSa-iso}
\end{align}
It can be compared with eq.~(6.45) of \cite{Levi:2015msa}.

\paragraph{Canonical transformation for NLO spin(a)-spin(b)}

For the spin(a)-spin(b) interaction term, we parametrize the Hamiltonian by 
\begin{align}
\begin{split}
H_{S_a S_b}^{\text{NLO}}  
= - \frac{1}{4} \left(\frac{G}{r^3}\right) 
& \left[h_1 p^2 (\vec{a}_b\cdot\vec{a}_b) + h_2 p^2 (\vec{a}_a\cdot\hat{n})  (\vec{a}_b\cdot\hat{n}) 
+ h_3 (\vec{p} \cdot \hat{n})^2 (\vec{a}_b\cdot\vec{a}_b)
\right.
\\
&\quad 
+ h_4 (\vec{p} \cdot \hat{n})^2 (\vec{a}_a\cdot\hat{n})  (\vec{a}_b\cdot\hat{n}) 
+ h_5 (\vec{p} \cdot \vec{a}_a)(\vec{p} \cdot \vec{a}_b) 
\\
&\quad\quad \left.
+ \frac{1}{2} (\vec{p} \cdot \hat{n}) \{ h_6 (\vec{p} \cdot \vec{a}_a) (\vec{a}_b \cdot \hat{n})  + \bar{h}_6 (\vec{p} \cdot \vec{a}_b) (\vec{a}_a\cdot\hat{n}) \}\right] \,,
\end{split}
\label{NLO-SSab-family}
\end{align}
and generators of transformation, 
\begin{align}
\begin{split}
g_{S_a S_b}^{\text{NLO}} = - \frac{m_a}{4(1+\zeta^{-1})} \left( \frac{G}{r^2}\right) 
&\left[ g_1 (\vec{p} \cdot \hat{n}) (\vec{a}_b\cdot\vec{a}_b) 
+ g_2  (\vec{p} \cdot \hat{n} ) (\vec{a}_a\cdot\hat{n})  (\vec{a}_b\cdot\hat{n}) 
 \right.
\\
& \quad \qquad 
\left.  
 + \frac{1}{2} \{ g_3 (\vec{p} \cdot \vec{a}_a) (\vec{a}_b \cdot \hat{n}) + \bar{g}_3 (\vec{p} \cdot \vec{a}_b) (\vec{a}_a\cdot\hat{n}) \}
 \right] \,.
\end{split}
\label{gen-SSab}
\end{align}

Our result \eqref{NLO-SSb-iso} amounts to 
\begin{align}
h_1 = 2 \zeta + 9 + 2 \zeta^{-1} \,,
\quad 
h_2 = -3 h_1 \,,
\quad 
h_3 = h_4 = h_5 = h_6 = 0 \,.
\end{align}
In eq.(6.10) of \cite{Levi:2014sba}, the $h$ parameters are
\begin{equation}
\begin{array}{llll}
h_1 = 6 \zeta + 16 +  6 \zeta^{-1} \,,
&
h_2 = -6 \zeta -21  -6 \zeta^{-1} \,,
&
h_3 = -21 \zeta -12  -12  \zeta^{-1} \,,
\\
h_4 = -30 \,, 
&
h_5 = -6 \zeta -14 -6  \zeta^{-1} \,,
&
h_6 = 12 \zeta + 54 +  24 \zeta^{-1} \,,
\end{array}
\end{equation}
and $\bar{h}_6 = h_6|_{\zeta \rightarrow 1/\zeta}$. 
Taking the difference $\Delta h_k = h_k^{\rm EFT} - h_k^{\rm amp}$, we find
\begin{equation}
\begin{array}{llll}
\Delta h_1 = 4 \zeta + 7 +  4 \zeta^{-1} \,,
&
\;\; \Delta h_2 = 6 \,,
&
\;\; \Delta h_3 = -21 \zeta -12  -12  \zeta^{-1} \,,
\\
\Delta h_4 = -30 \,, 
&
\;\; \Delta h_5 = -6 \zeta -14 -6  \zeta^{-1} \,,
&
\;\; \Delta h_6 = 12 \zeta + 54 +  24 \zeta^{-1} \,,
\end{array}
\label{SSz}
\end{equation}
The canonical transformation at NLO relates $\Delta h_k$ to $g_k$ as
\begin{equation}
\begin{array}{lll}
\Delta h_1 = -g_1 \,,
&
\quad \Delta h_2 = -g_2 \,,
&
\quad \Delta h_3 = 3 g_1 \,,
\\
\Delta h_4 = 5 g_2 \,, 
&
\quad \Delta h_5 =- (g_3 + \bar{g}_3)/2 \,,
&
\quad \Delta h_6 = 3g_3 - 2g_2 \,.
\end{array}
\label{SSw}
\end{equation}
The differences \eqref{SSz} match the relations \eqref{SSw} precisely if we set 
\begin{equation}\label{eq:1408NLO matching}
\begin{array}{llll}
g_1 = -(4 \zeta + 7 +  4 \zeta^{-1}) \,,
&
\quad g_2 = -6 \,,
&
\quad g_3 = 4 \zeta + 14  + 8  \zeta^{-1} \,,
&
\bar{g}_3 = g_3|_{\zeta \rightarrow 1/\zeta}\,.
\end{array}
\end{equation}

\paragraph{Canonical transformation for NLO spin(a)-squared} 

For the spin(a)-squared term, we consider a family of Hamiltonians: 
\begin{align}
\begin{split}
H_{S_a^2}^{\rm NLO} = - \frac{1}{8} \left(\frac{G}{r^3}\right) 
& \left[h_1 \vec{p}^2 \vec{a}_a^2  + h_2 p^2 (\vec{a}_a \cdot \hat{n})^2 
+ h_3 (\vec{p} \cdot \hat{n})^2 \vec{a}_a^2 \right.
\\
&\quad \left.
+ h_4 (\vec{p} \cdot \hat{n})^2 (\vec{a}_a \cdot \hat{n})^2
+ h_5 (\vec{p} \cdot \vec{a}_a)^2 + h_6  (\vec{p} \cdot \hat{n}) (\vec{p} \cdot \vec{a}_a) (\vec{a}_a \cdot \hat{n})  \right] \,,
\end{split}
\label{NLO-SS-family}
\end{align}
and generators of transformation, 
\begin{align}
g_{S_a^2}^{\rm NLO} &= - \frac{m_a}{8(1+\zeta^{-1})} \left( \frac{G}{r^2}\right) 
\left[ g_1 (\vec{p} \cdot \hat{n}) \vec{a}_a^2 + g_2 (\vec{p} \cdot \hat{n} ) (\vec{a}_a \cdot \hat{n})^2  + g_3 (\vec{p} \cdot \vec{a}_a) (\vec{a}_a \cdot \hat{n})  \right] \,.
\label{gen-SS}
\end{align}
 
Our result \eqref{NLO-SSa-iso} amounts to $(C_{\rm here} = C_2^{(a)}$)
\begin{align}
h_1 = (6 \zeta + 16 + 6 \zeta^{-1})C - (7\zeta +8) \,,
\quad 
h_2 = -3 h_1 \,,
\quad 
h_3 = h_4 = h_5 = h_6 = 0 \,.
\end{align}
This is to be compared with eq.(6.45) of \cite{Levi:2015msa}. 
Reducing it to the COM frame, we obtain a somewhat simplified formula in our notation,
\begin{equation}
\begin{array}{ll}
h_1 = (10\zeta + 18 + 6 \zeta^{-1} )C - (10\zeta + 12)  \,, 
&
h_2 = - (18\zeta + 42 + 18\zeta^{-1} ) C + 21 \zeta + 24  \,,
\\
h_3 =  -(12 \zeta +6) C + 9 \zeta + 12 \,, 
&
h_4 = -30C \,, 
\\
h_5 = -(4\zeta + 4) C + 10\zeta + 12 \,, 
&
h_6 = (12 \zeta + 24) C - (30\zeta + 36) \,. 
\end{array}
\end{equation}
Taking the difference, $\Delta h_k = h_k^{\rm EFT} - h_k^{\rm amp}$, we find
\begin{equation}
\begin{array}{ll}
\Delta h_1 = (4\zeta +2) C - (3 \zeta + 4) \,, 
&
\quad \Delta h_2 = 6 C \,,
\\
\Delta h_3 = -(12 \zeta +6) C + 9 \zeta + 12 \,,
&
\quad \Delta h_4 = -30C \,, 
\\
\Delta h_5 = -(4\zeta + 4) C + 10\zeta + 12 \,,
&
\quad \Delta h_6 = (12 \zeta + 24) C - (30\zeta + 36) \,.
\end{array}
\label{SSx}
\end{equation}
Performing the canonical transformation, we relate $\Delta h_k$ to $g_k$:
\begin{equation}
\begin{array}{ll}
\Delta h_1 = -g_1 \,,
&
\quad \Delta h_2 = -g_2 \,,
\\
\Delta h_3 = 3 g_1 \,,
&
\quad \Delta h_4 = 5 g_2 \,, 
\\
\Delta h_5 = - g_3 \,,
&
\quad \Delta h_6 = 3g_3 - 2g_2 \,.
\end{array}
\label{SSy}
\end{equation}
The differences \eqref{SSx} match the relations \eqref{SSy} precisely, if we set 
\begin{align}
g_1 = -(4\zeta +2) C + (3 \zeta + 4) \,,
\quad
g_2 = - 6C \,,
\quad 
g_3 = (4\zeta + 4) C - (10\zeta + 12) \,.
\end{align}

\paragraph{NNLO}

The NNLO spin-spin term is
\begin{align}
\begin{split}
X^{\rm NNLO}_{(1,1)} &= 
\left( 
\frac{ - 6 m_a^4 -15 m_a^3 m_b^2 + 4 m_a^2 m_b^2  -15 m_a m_b^3 -6 m_b^4 }{16m_a^4 m_b^4}  \right) \vec{p}^4 \,.
\end{split}
\label{NNLO-SSab}
\end{align}
The NNLO spin-squared term is
\begin{align}
\begin{split}
X^{\rm NNLO}_{(2,0)} &= \left( \frac{ C_2^{(a)} (-5m_a^4 + 18m_a^2m_b^2 -5m_b^4) - 3(7m_a^2 +4 m_a m_b -2 m_b^2) m_b^2 }{16m_a^4m_b^4} \right) \vec{p}^4\,.
\end{split}
\label{NNLO-SSa-iso}
\end{align}
These are to be compared with eqs.(3.3)-(3.4)  of \cite{Levi:2016ofk}.

\paragraph{Canonical transformation for NNLO  spin(a)-spin(b)}

We parametrize the spin(a)-spin(b) term of the Hamiltonian at the NNLO order as
\begin{equation}
\begin{split}
H_{S_a S_b}^{\text{NNLO}} 
=&
\frac{1}{16m_a m_b}
\left( \frac{G}{r^3} \right)
\Big[
(\hat{n}\cdot\vec{p})^4 
\left[ h_7 (\vec{a}_a\cdot\vec{a}_b) + h_8 (\hat{n}\cdot \vec{a}_a)(\hat{n}\cdot \vec{a}_b) \right]\\
&+
\frac{1}{2}(\hat{n}\cdot\vec{p})^3 
\left[ h_9 (\hat{n}\cdot\vec{a}_a)(\vec{p}\cdot\vec{a}_b) + \bar{h}_9 (\hat{n}\cdot\vec{a}_b)(\vec{p}\cdot\vec{a}_a) \right]\\
&+
(\hat{n}\cdot\vec{p})^2
\left[h_{10} (\vec{p}\cdot\vec{a}_a)(\vec{p}\cdot\vec{a}_b) + h_{11} \vec{p}^2\vec{a}_a\cdot\vec{a}_b +  h_{12}\vec{p}^2(\hat{n}\cdot \vec{a}_a)(\hat{n}\cdot \vec{a}_b)\right]\\
&+
\frac{1}{2}(\hat{n}\cdot\vec{p})
\left[ h_{13}\vec{p}^2(\hat{n}\cdot\vec{a}_a)(\vec{p}\cdot\vec{a}_b) + \bar{h}_{13}\vec{p}^2(\hat{n}\cdot\vec{a}_b)(\vec{p}\cdot\vec{a}_a) \right]\\
&+
h_{14}\vec{p}^2 (\vec{p}\cdot\vec{a}_a)(\vec{p}\cdot\vec{a}_b) 
+
\vec{p}^4 \left[ h_{15} (\vec{a}_a\cdot\vec{a}_b) + h_{16} (\hat{n}\cdot \vec{a}_a)(\hat{n}\cdot \vec{a}_b) \right]
\Big] \,.
\end{split}
\end{equation}
The NNLO generator is parametrized as
\begin{equation}
\begin{split}
g_{S_a S_b}^{\text{NNLO}} = -& \frac{1}{16m_b(1+\zeta^{-1})}  \left( \frac{G}{r^2}\right) \Big[
(\hat{n}\cdot\vec{p})^3 \left[g_4 (\vec{a}_a\cdot\vec{a}_b)+g_5 (\hat{n}\cdot \vec{a}_a)(\hat{n}\cdot \vec{a}_b)\right] \\
+&
(\hat{n}\cdot\vec{p})^2 \left[ g_6 (\hat{n}\cdot \vec{a}_a) (\vec{p}\cdot \vec{a}_b) + \bar{g}_6 (\hat{n}\cdot \vec{a}_b) (\vec{p}\cdot \vec{a}_a) \right]\\
+&
(\hat{n}\cdot\vec{p}) \left[g_7 (\vec{p}\cdot\vec{a}_a)(\vec{p}\cdot\vec{a}_b) + g_8\vec{p}^2 (\hat{n}\cdot \vec{a}_a)(\hat{n}\cdot \vec{b}_a) + g_9\vec{p}^2 \vec{a}_a \cdot \vec{a}_b\right] \\
+& \vec{p}^2\left[ g_{10} (\hat{n}\cdot \vec{a}_a) (\vec{p}\cdot \vec{a}_b) + \bar{g}_{10} (\hat{n}\cdot \vec{a}_b) (\vec{p}\cdot \vec{a}_a) \right]
\Big] \,.
\end{split}
\end{equation}

Our result \eqref{NNLO-SSab} amounts to 
\begin{equation}
\begin{split}
&
h_7 = h_8 = h_9 = \bar{h}_9 = h_{10} = h_{11} = h_{12} = h_{13} = \bar{h}_{13} = h_{14} = 0, 
\\
&
h_{15} = 6 \zeta ^2 +15 \zeta-4 +15 \zeta^{-1} +6\zeta ^{-2},
\quad 
h_{16} = -3 h_{15}.
\end{split}
\label{SS-NNLO-amp}
\end{equation}
Eq.(6.12) of \cite{Levi:2014sba}, translated to our notation, yields
\begin{equation}
\begin{split}
&
h_7 =  0, \quad
h_8 = 210, \quad
h_9 =  -60 ( 5 + 2\zeta^{-1}), \quad 
\bar{h}_9 = h_9|_{\zeta \rightarrow \zeta^{-1}} \,,
\\
&
h_{10} = 12 \left(\zeta + 5 + \zeta^{-1} \right), \quad
h_{11} =- 3 \left(16 \zeta ^2 + 19 \zeta + 14 +19 \zeta^{-1}  +16 \zeta^{-2} \right), \quad
\\
&
h_{12} = 90, \quad
h_{13} = 6 \left(16 \zeta ^2 +19 \zeta +6 +19 \zeta^{-1}  +6 \zeta^{-2}\right), 
\quad
\bar{h}_{13} = h_{13}|_{\zeta \rightarrow \zeta^{-1}} \,,
\\
&
h_{14} = -2(11 \zeta ^2 + 15 \zeta + 4  + 15\zeta^{-1} \zeta + 11\zeta^{-2}), 
\\
&
h_{15} = 22 \zeta ^2 +34 \zeta +10 +34 \zeta^{-1} +22 \zeta^{-2} \,,\\
&
h_{16} = -3\left(6 \zeta ^2 + 15 \zeta + 8 + 15 \zeta^{-1} +6 \zeta^{-2}\right) \,.
\end{split} 
\label{SS-NNLO-EFT}
\end{equation}
The changes in the $h$ parameters are related by \eqref{cano-NNLO} to the $g$ parameters as
\begin{equation}
\begin{split}
&
\Delta h_7 =  - 5g_4, \quad 
\Delta h_8 = - 7g_5, 
\\
&
\Delta h_9 =  2(g_5 - 5g_6), \quad
\Delta \bar{h}_9 =    2(g_5 - 5\bar{g}_6), \quad
\Delta h_{10} = g_6 +\bar{g}_6 - 3g_7, \\
&
\Delta h_{11} =  6\left(\zeta - 1 + \zeta^{-1} \right) g_1 + 3 \left(g_4-g_9\right),
\\
&
\Delta h_{12} =  10 \left(\zeta - 1 + \zeta^{-1} \right) g_2 + (3g_5-5g_8), \\
&
\Delta h_{13} =  2(2 g_6+g_8-3 g_{10}) - 2 \left(\zeta - 1 + \zeta^{-1} \right) \left(2 g_2-3 \bar{g}_3\right),
\\
&
\Delta \bar{h}_{13} = 2(2 \bar{g}_6+g_8-3 \bar{g}_{10} ) - 2 \left(\zeta - 1 + \zeta^{-1} \right) \left(2 g_2-3 g_3\right), \\
&
\Delta h_{14} =   g_7 +g_{10}+ \bar{g}_{10} - \left(\zeta - 1 + \zeta^{-1} \right) \left(g_3+ \bar{g}_{3}\right), 
\\
&
\Delta h_{15} = g_9- 2 \left(\zeta - 1 + \zeta^{-1} \right) g_1,  \quad
\Delta h_{16} =  g_8 - 2\left(\zeta - 1 + \zeta^{-1} \right)  g_2 .
\end{split}
\end{equation}
The difference $\Delta h = h^{\rm EFT} - h^{\rm amp}$ between 
\eqref{SS-NNLO-EFT} and \eqref{SS-NNLO-amp} is accounted for if we choose the $g$ parameters as 
\begin{equation}
\begin{split}
&
g_4 = 0, \quad 
g_5 = -30, \quad 
g_6 = 12 (2 + \zeta^{-1}) , \quad 
\bar{g}_6 = 12 (\zeta +2), \\
& 
g_7 = -4,\quad
g_8 = - 12 \left(\zeta + 2 + \zeta^{-1} \right), \quad 
g_9 = 8 \zeta ^2 + 13 \zeta + 12  + 13 \zeta^{-1} +8\zeta^{-2}, \quad \\
&
g_{10} = -(8 \zeta^2 + 13 \zeta +4 + \zeta^{-1} +2\zeta^{-2}), \quad
\bar{g}_{10} = -(2 \zeta ^2 + \zeta + 4+13 \zeta^{-1} +8\zeta^{-2}).
\end{split}
\end{equation}


\paragraph{Canonical transformation for NNLO spin(a)-squared} 

We parametrize  the NNLO spin(a)$^2$ sector Hamiltonian as
\begin{equation}
\begin{split}
 H_{S_a^2}^{\text{NNLO}}
= \frac{1}{16 m_a m_b r^3}  \left( \frac{G}{r^3}\right) & \Big[
(\hat{n}\cdot\vec{p})^4 
\left[ 
h_7\vec{a}_a^2 + 
h_8 (\hat{n}\cdot \vec{a}_a)^2 \right] \\
&+
(\hat{n}\cdot\vec{p})^3 
\left[ h_9(\hat{n}\cdot\vec{a}_a)(\vec{p}\cdot\vec{a}_b)  \right]\\
&+
(\hat{n}\cdot\vec{p})^2
\left[ 
h_{10}(\vec{p}\cdot\vec{a}_a)^2 
+h_{11} \vec{p}^2\vec{a}_a^2
+ h_{12} \vec{p}^2(\hat{n}\cdot \vec{a}_a)^2
\right]\\
&+
(\hat{n}\cdot\vec{p})
\left[ h_{13}\vec{p}^2(\hat{n}\cdot\vec{a}_a)(\vec{p}\cdot\vec{a}_b)  \right]\\
&+
h_{14}\vec{p}^2 (\vec{p}\cdot\vec{a}_a)^2 
+ 
\vec{p}^4 \left[h_{15} \vec{a}_a^2 +  h_{16}(\hat{n}\cdot \vec{a}_a)^2 \right]
\Big] \,.
\end{split}
\end{equation}
The NNLO generator is parametrized as
\begin{equation}
\begin{split}
g_{S_a^2}^{\text{NNLO}} = - \frac{1}{16m_b (1+\zeta^{-1})} \left( \frac{G}{r^2}\right)  \Big[
&(\hat{n}\cdot\vec{p})^3 \left[g_4 \vec{a}_a^2+g_5 (\hat{n}\cdot \vec{a}_a)^2\right] \\
+&
(\hat{n}\cdot\vec{p})^2 \left[ g_6 (\hat{n}\cdot \vec{a}_a) (\vec{p}\cdot \vec{a}_a)  \right]\\
+&
(\hat{n}\cdot\vec{p}) \left[g_7 (\vec{p}\cdot\vec{a}_a)^2 + g_8\vec{p}^2 (\hat{n}\cdot \vec{a}_a)^2+ g_9\vec{p}^2 \vec{a}_a^2 \right] \\
+& \vec{p}^2\left[ g_{10} (\hat{n}\cdot \vec{a}_a)(\vec{p} \cdot \vec{a}_a)   \right]
\Big]
\end{split}
\end{equation}
The changes in the $h$ parameters are related by \eqref{cano-NNLO} to the $g$ parameters as
\begin{equation}
\begin{split}
&
\Delta h_7 =  - 5g_4, \quad 
\Delta h_8 = - 7g_5 , \quad
\Delta h_9 =  2g_5 - 5g_6 , \quad
\Delta h_{10} =   g_6  - 3g_7 , \\
&
\Delta h_{11} =  3\left(\zeta -1 + \zeta^{-1} \right) g_1 + 3 \left(g_4-g_{9}\right),\quad
\Delta h_{12} =  5\left(\zeta - 1 + \zeta^{-1} \right) g_2 + 3g_5-5g_8 , \\
&
\Delta h_{13} =  2 g_6+2g_8-3 g_{10} - \left(\zeta - 1 + \zeta^{-1} \right) \left(2 g_2-3 g_3\right), \\
&
\Delta h_{14} = g_7+g_{10} - \left(\zeta - 1 + \zeta^{-1} \right) g_3, \quad
\Delta h_{15} = g_9 - \left(\zeta - 1 + \zeta^{-1} \right) g_1 , \\
&
\Delta h_{16} =  g_8 - \left(\zeta - 1 + \zeta^{-1} \right) g_2 .
\end{split}
\end{equation}

Our result \eqref{NNLO-SSa-iso} amounts to 
\begin{equation}
\begin{split}
&
h_7 = h_8 = h_9 = h_{10} = h_{11} = h_{12} = h_{13} = h_{14} = 0 ,\\
&
h_{15} = C \left(5 \zeta ^2 -18 + 5  \zeta^{-2} \right)+ (-6 \zeta ^2+12 \zeta +21 ),
\quad 
h_{16} = -3 h_{15} .
\end{split}
\label{SSaa-NNLO-amp}
\end{equation}
This can be compared with eq.(3.4) of \cite{Levi:2016ofk} 
which adopts the same coordinate gauge as \cite{Levi:2015msa}: 
\begin{align}
\begin{split}
&
h_7 = 15\left( \zeta+ 4 + 4 \zeta^{-1}\right) + 15C(3+4\zeta), \quad 
h_8 = 105 C, \\
&
h _9 = - 30 C(4 + 2\zeta) -30(\zeta + 6 + 6\zeta^{-1}) , \\
&
h_{10} =  6(-2\zeta^2-4\zeta+6+10\zeta^{-1}) + 6C(\zeta^2 + 6\zeta + 6) , \\
&
h_{11} = 6C (2\zeta^2 + 6 \zeta + 5 + 2\zeta^{-1}) + 3(5\zeta^2 - 6 \zeta - 33 - 24\zeta^{-1} ) ,\\
&
h_{12} = 30C(2\zeta + 5 + 2\zeta^{-1}) + 15(\zeta + 8 + 8\zeta^{-1}) , \\
&
h_{13} = -6C(3\zeta^2 + 18 \zeta+26+8\zeta^{-1}) - 3(7\zeta^2-27\zeta-50-12\zeta^{-1}) , \\
&
h_{14} = 4C(\zeta^2+ 7 \zeta + 8 + 2\zeta^{-1})+(11\zeta^2-15\zeta-42-12\zeta^{-1}) , \\
&
h_{15} = C(\zeta^2 - 24\zeta-37-4\zeta^{-1}+5\zeta^{-2}) + (-11\zeta^2 + 15 \zeta + 42 + 12\zeta^{-1}) , \\
&
h_{16} = -3C (5\zeta^2 + 4 \zeta -5 + 4\zeta^{-1}+5\zeta^{-2}) - 3(-6\zeta^2 + 13 \zeta + 29 + 8 \zeta^{-1}) .\\
\end{split}
\label{SSaa-NNLO-EFT}
\end{align}
The difference $\Delta h = h^{\rm EFT} - h^{\rm amp}$ between 
\eqref{SSaa-NNLO-EFT} and \eqref{SSaa-NNLO-amp} is accounted for if we choose the $g$ parameters as
\begin{equation}
\begin{split}
&
g_4 = - 3(3+4\zeta)C - 3\left(4 \zeta^{-1} +4 +\zeta \right), \quad 
g_5 = - 15 C, \\
&
g_6 = 6 (2 \zeta +3) C + 2 (3\zeta + 18 + 18 \zeta^{-1}), \quad \\
& 
g_7 = -2C(\zeta^2 + 4\zeta + 3) - 2\left(-2\zeta^2 - 5 \zeta + 4 \zeta^{-1}\right) , \quad\\
&
g_8 = - 3(6\zeta + 11 + 6\zeta^{-1}) C - 3\left(\zeta + 8 + 8 \zeta^{-1}\right) , \quad  \\
&
g_9 = -(8 \zeta^2 + 22\zeta + 21 + 6\zeta^{-1} ) C + 2( -\zeta^2 + 2 \zeta +10  +8 \zeta^{-1} ) , \quad \\
&
g_{10} = 2 (1 + \zeta^{-1}) (\zeta +2) (5 \zeta +3) C + (1 + \zeta^{-1}) (-3 \zeta^2-24 \zeta -16)  .
\end{split}
\end{equation}


\subsection{Cubic in spin (up to NLO)}

\paragraph{Cubic in spin} 

The mixed cubic term is given by 
\begin{align}
X_{(2,1)}^{\rm NLO} = 3 \left( \frac{C_2^{(a)} m_a (-5 m_a^2 + 8 m_a m_b + 18 m_b^2) - (4 m_a^2 + 21 m_a m_b + 6 m_b^2) m_b }{16m_a^2 m_b^3 } \right) \vec{p}^2
 \,.
\end{align}
The self-cubic term is the first term to host the ``magnetic" Wilson coefficient $C_3$:
\begin{align}
X_{(3,0)}^{\rm NLO} &= \left( \frac{24 C_3^{(a)} m_a  (m_a + m_b) 
- 3 C_2^{(a)} (6 m_a^2 + 16 m_a m_b + 5 m_b^2) 
+  (12m_a +11 m_b)m_b}{16m_a^2 m_b^2} \right)\vec{p}^2 \,.
\end{align} 
These are to be compared with the 1PM parts of the recent PN computation given in \cite{Levi:2019kgk}. 
But, since the result of \cite{Levi:2019kgk} in its current form carry time derivatives of momenta, 
which are not visible in our framework, a direct comparison requires more than a canonical transformation. We leave the comparison for a future work. 
%

\section{Effective one-body mapping} \label{sec:EOB}

Newtonian mechanics with translation symmetry exhibits a complete decoupling of the center-of-mass coordinates from the relative coordinates. Such a decoupling is obscure in GR. 
In perturbative approaches to GR, the EOB mapping offers a 
way to map a binary system to a test-body in the background of a massive ``center". 

The EOB mapping was originally introduced in a PN context \cite{Buonanno:1998gg}, and a PM version was introduced more recently in \cite{Damour:2016gwp,Damour:2017zjx}. For a Kerr black hole black hole background, Vines \cite{Vines:2017hyw} carried out the PM-EOB mapping to the 1PM order and to all orders in spin. 

In this section, we show that our 1PM Hamiltonian restricted to Kerr black holes agree perfectly 
with Vines' result to the 1PM order in the EOB mapping. 
We do not provide a proper review of the EOB theory and refer the readers to 
the original papers.

\paragraph{Spin-less test body} 

The Schwartzschild metric in the isotropic coordinate is 
\begin{align}
ds^2 = \left(1 - \frac{GM}{2r} \right)^2  \left(1 + \frac{GM}{2r} \right)^{-2} dt^2 
- \left(1 + \frac{GM}{2r} \right)^{4} d\vec{x}^2 \,.
\end{align}
To the 1PM order, it is fine to approximate the metric by 
\begin{align}
ds^2 \approx \left(1 - \frac{2GM}{r} \right)dt^2 
- \left(1 + \frac{2GM}{r} \right) d\vec{x}^2 \,.
\end{align}
The Lagrangian of a test-body moving in this background is (subscript t for `test body')
\begin{align}
L_{\rm t} = - \mu \sqrt{ \left(1 - \frac{2GM}{r} \right)
- \left(1 + \frac{2GM}{r} \right) \left(\frac{d\vec{x}}{dt}\right)^2 } \,.
\end{align}
The corresponding Hamiltonian is 
\begin{align}
H_{\rm t} =  \sqrt{ \left(1 - \frac{2GM}{r} \right)
\left( \mu^2 +  \left(1 - \frac{2GM}{r} \right) \vec{p}_{\rm t}^2 \right)} \,.
\end{align}
Truncating to the 1PM order, we find an expression exact in $\vec{p}_{\rm t}^2$:
\begin{align}
(H_{\rm t})_{\rm 1PM} =  \mu \left[ \gamma - \frac{GM}{r}  \left( \frac{2\gamma^2 -1}{\gamma}\right) \right]\,, 
\quad 
\gamma = \sqrt{1+ \vec{p}_{\rm t}^2/\mu^2} \,.
\label{test-c1}
\end{align}

This should {\em not} be compared directly with the 1PM two-body Hamiltonian (subscript r for `real'):
\begin{align}
H_{\rm r} = E_a + E_b - \frac{Gm_am_b}{r}\left(\frac{m_a m_b}{E_a E_b}\right) (2\sigma^2 -1)  \,,
\label{real-c1}
\end{align}
where we recall 
\begin{align}
E_{a/b} = \sqrt{ m^2_{a/b} + \vec{p}^2} \,,
\quad 
\sigma = \frac{p_a \cdot p_b}{m_a m_b} = \frac{E_a E_b + \vec{p}^2}{m_a m_b} \,.
\end{align}

One way to motivate the correct EOB mapping is to match the deflection angle $\chi$ of a scattering process. 
As explained in \cite{Damour:2016gwp}, the EOB mapping requires that 
\begin{align}
\chi_{\rm real}(E_{\rm real} , J ) = \chi_{\rm test}(E_{\rm test}, J)\,, 
\label{deflection}
\end{align}
where $J$ is the angular momentum, and the map between $E_{\rm real}$ and $E_{\rm test}$ can be non-trivial. 

\paragraph{Quasi-Newtonian approach}

Recent papers \cite{Kalin:2019rwq,Bjerrum-Bohr:2019kec,Damour:2019lcq} suggest a way to understand the EOB mapping before computing the deflection angle. The key idea is to invert the energy-momentum relation of any PM-like theory, 
\begin{align}
E = H(p,r) = c_0(p) + \sum_{n=1}^\infty \left(\frac{G}{r} \right)^n c_n(p)
\end{align}
to find an expression of the form 
\begin{align}
p^2 = p_\infty^2 + \sum_{n=1}^\infty \left(\frac{G}{r} \right)^n w_n(p_\infty) \,, 
\end{align}
where the gauge-invariant asymptotic momentum, $p_\infty$ is defined implicitly by 
\begin{align}
E = H(p_\infty, r=\infty)\,.
\end{align}
If two systems yield the same quasi-Newtonian functions $w_n(p)$ to all orders, 
it is guaranteed that the two systems agree on the deflection angle.
To the linear order in $G$, it is easy to do the inversion 
and verify the `quasi-Newtonian duality'. 

In the two-body picture,
\begin{align}
\begin{split}
E_{\rm r} &= \sqrt{m_a^2 + p_\infty^2} + \sqrt{m_b^2 + p_\infty^2} 
\\
&= \sqrt{m_a^2 + p^2} + \sqrt{m_b^2 + p^2}  + \left(\frac{G}{r}\right) c_1(p)  
\\
&= \sqrt{m_a^2 + p_\infty^2 + (G/r) w_1} + \sqrt{m_b^2 + p_\infty^2 + (G/r)w_1} 
+ \left(\frac{G}{r}\right) c_1(p_\infty) + \mathcal{O}(G^2) 
\,.
\end{split}
\end{align}
Expanding the square-roots and demanding that the $\mathcal{O}(G)$ terms cancel out, we find 
\begin{align}
w_1(p_\infty) = -\frac{2E_a E_b}{E} c_1(p_\infty) \,.
\end{align}
In the effective one-body picture,
\begin{align}
\begin{split}
E_{\rm t} = \sqrt{\mu^2 + p_\infty^2} 
&= \sqrt{m_a^2 + p^2}  + \left(\frac{G}{r}\right) \tilde{c}_1(p)  
\\
&= \sqrt{m_a^2 + p_\infty^2 + (G/r)\widetilde{w}_1 } 
+ \left(\frac{G}{r}\right) \tilde{c}_1(p_\infty) + \mathcal{O}(G^2) 
\,.
\end{split}
\end{align}
It follows that 
\begin{align}
\widetilde{w}_1(p_\infty) = - (2E_{\rm t})\tilde{c}_1(p_\infty) \,.
\end{align}

Reading off $c_1$ from \eqref{real-c1} and $\tilde{c}_1$ from \eqref{test-c1}, we find 
\begin{align}
w_1 = 2 \frac{m_a^2 m_b^2}{E} (2\sigma^2 -1) \,,
\quad 
\widetilde{w}_1 = 2 E_t M \mu \frac{2\gamma^2 -1}{\gamma} = 2M \mu^2 (2\gamma^2-1)\,.
\end{align}
Assuming the 0PM EOB dictionary, 
\begin{align}
M = m_a + m_b \,,
\quad 
\mu = \frac{m_a m_b}{m_a +m_b} \,,
\quad 
\sigma = \gamma\,,
\end{align}
we find 
\begin{align}
\frac{\widetilde{w}_1}{w_1} = \frac{E}{M} \equiv \Gamma\,.
\end{align}
This ratio can be absorbed by rescaling the position and momenta variables 
as follows:
\begin{align}
(\vec{p})_{\rm t} = \Gamma \, \vec{p}_{\rm r} 
\quad 
(\vec{r})_{\rm t} = \Gamma^{-1} \vec{r}_{\rm r}  \,.
\label{EOB-rescaling}
\end{align}
This scaling is consistent with the invariance of the angular momentum 
assumed in \eqref{deflection}.

\paragraph{Vines' spinning EOB} 

Vines \cite{Vines:2017hyw} considered a spinning test-body 
in the Kerr black hole background and arrived at 
\begin{align}
\frac{H_{\rm t}}{\mu} = \gamma - \frac{GM}{2\gamma} 
\sum_{s=\pm 1} 
(\gamma + s \sqrt{\gamma^2-1})^2
\left| \boldsymbol{R} +\frac{\boldsymbol{P}}{\mu} \times \left(s\frac{\boldsymbol{a}_{\rm b} +\boldsymbol{a}_{\rm t} }{\sqrt{\gamma^2-1}} - \frac{\boldsymbol{a}_{\rm t}}{\gamma+1} \right) 
\right|^{-1} \,,
\end{align}
where ($\boldsymbol{R}$, $\boldsymbol{P}$) translate to $(\vec{r}, \vec{p})_{\rm test}$ in our notation, and ($\boldsymbol{a}_{\rm b}$, $\boldsymbol{a}_{\rm t}$) 
are spin-length vectors to be explained shortly. 
According to the EOB dictionary we have mentioned, including the rescaling \eqref{EOB-rescaling}, this test-body Hamiltonian corresponds to the 1PM two-body potential of the form 
\begin{align}
V = - \left( \frac{G}{r}\right) \frac{m_a^2m_b^2}{2E_a E_b} 
\sum_{s=\pm 1} 
e^{2s\theta}
\left| \vec{r} +\frac{\Gamma^2}{\mu} \vec{p} \times \left(s\frac{\boldsymbol{a}_{\rm b} +\boldsymbol{a}_{\rm t} }{\sinh\theta} - \frac{\boldsymbol{a}_{\rm t}}{ \cosh\theta +1} \right) 
\right|^{-1} \,. 
\label{EOB-almost}
\end{align}
The relation between ($\boldsymbol{a}_{\rm b}$, $\boldsymbol{a}_{\rm t}$) and  
$(\vec{a}_a,\vec{a}_b)$ in this paper was already discussed in \cite{Vines:2017hyw}:
\begin{align}
\boldsymbol{a}_{\rm b} + \boldsymbol{a}_{\rm t} = \frac{1}{\Gamma} \vec{a}_0 \,,
\quad 
\frac{\boldsymbol{a}_{\rm t}}{\mu ( \cosh\theta +1)} = \frac{1}{\Gamma^2} \left( \frac{\vec{a}_a}{E_a + m_a } +\frac{\vec{a}_b}{E_b + m_b } \right) \,.
\end{align}
Substituting these into \eqref{EOB-almost}, we finally obtain 
\begin{align}
V = - \left( \frac{G}{r}\right) \frac{m_a^2m_b^2}{2E_a E_b} 
\sum_{s=\pm 1} 
e^{2s\theta}
\left| \vec{r} + s\frac{E (\vec{p}\times \vec{a}_0)}{m_a m_b \sinh\theta} - \frac{\vec{p}\times \vec{a}_a}{m_a r_a} - \frac{\vec{p}\times \vec{a}_b}{m_b r_b} 
\right|^{-1} \,. 
\nonumber
\end{align}
It agrees perfectly with our 1PM potential for Kerr black holes shown in \eqref{V-BH-final}.

\section{Conclusion and outlook} \label{sec:conclusion}

In this paper we derived the exact 1PM gravitational potential to all orders in spin for a binary of general compact spinning bodies. Building on previous work~\cite{Chung:2019duq}, we derived the exact Thomas-Wigner rotation factor, which allows us to construct the exact 1PM potential. Through highly non-trivial canonical transformations, match to existing PN results were made as well as the effective one-body mapping for Kerr black holes. Our result exhibits almost complete factorization between the spin- and velocity-dependent terms, which is absent in the previous approaches due to ubiquitous $S\cdot p$ terms. From our on-shell point of view, this is simply a reflection of classical dynamics being encoded in the factorization limit of the elastic scattering, which must factorize from unitarity. Since the general three-point amplitude is expressed as a multipole of $S\cdot q$, coupled with the form of the Thomas-Wigner rotation factor, this immediately leads to the absence of $S\cdot p$ terms. Thus the on-shell approach naturally exposes hidden simplicity for the classical potential. 

We expect the same simplicity to persist at 2PM since the exact Thomas-Wigner rotation factor is independent of $S\cdot p$, and the gravitational Compton amplitude does not generate such terms when inserted into the discontinuity associated with the non-analyticity for one-loop amplitudes~\cite{Chung:2019duq}. Proper subtraction for the iteration terms as well as improved understanding of higher-spin Compton amplitude is necessary for reaching a concrete statement.  

An interesting observation is that if one were ignorant of the Thomas-Wigner rotation factor, and directly extracting the potential from $M$, the result will match with previous EFT computations \textit{sans} terms that have time derivatives, such as $\dot{S}, \dot{v}$. In the EFT approach, these terms are removed via redefinition of variables. This last step then simply matches to our Thomas-Wigner rotation factor. An appropriate understanding of the physics behind this phenomenon is obviously desirable.    

Some recent papers \cite{Kalin:2019rwq,Bjerrum-Bohr:2019kec,Damour:2019lcq} (see also \cite{Guevara:2019fsj,Kosower:2018adc}) 
advocate approaches that extract gauge invariant observables, 
such as the deflection angle and the periastron advance, directly from scattering amplitudes 
without taking the PM or PN Hamiltonian as an intermediate step. 
It would be interesting to incorporate binary of general compact spinning bodies in these approaches.

\vskip 1cm 

\acknowledgments 

We are grateful for Michel Levi, Rodolfo Russo, Gabriele Travaglini and Congkao Wen for discussions. 
SL is grateful to the string theory groups at Queen Mary University of London and University of Iceland for hospitality where parts of this work were written.  
The work of JWK and SL was supported in part by the National Research Foundation of Korea grant NRF-2019R1A2C2084608.
MZC and YTH is supported by MoST Grant No. 106-2628-M-002-012-MY3. YTH is also supported by Golden
Jade fellowship.



\end{document}